\documentclass[a4paper,10pt]{article}
\usepackage[english]{babel}
\usepackage{jheppub}
\usepackage{subfig}
\usepackage{float}
\usepackage{caption}
\usepackage[T1]{fontenc} 
\usepackage{graphicx}
\usepackage{epsfig}
\usepackage{amsmath}
\usepackage{amssymb}
\usepackage{float}
\usepackage{placeins}
\usepackage{braket}
\usepackage{slashed}
\usepackage{mathdots}
\usepackage{lipsum}
\usepackage{bigints}
\usepackage{cleveref}

\usepackage{array}

\DeclareMathAlphabet\mathbfcal{OMS}{cmsy}{b}{n}

\allowdisplaybreaks[4]

\preprint{{
\begin{tabular}{l} 
  BONN-TH-2022-23\\ CERN-TH-2022-187\\ TTP22-068
\end{tabular}
}}

\title{Two-loop form factors for pseudo-scalar quarkonium production and decay}

\author[a,b]{Samuel Abreu,}
\author[c]{Matteo Becchetti,}
\author[d]{Claude Duhr,}
\author[e,f]{Melih~A.~Ozcelik}
\affiliation[a]{CERN, Theoretical Physics Department, CH-1211 Geneva 23, Switzerland}
\affiliation[b]{Higgs Centre for Theoretical Physics, School of Physics and Astronomy,
The University of Edinburgh, Edinburgh EH9 3FD, Scotland, UK}
\affiliation[c]{Physics Department, Torino University and INFN Torino,
Via Pietro Giuria 1, I-10125 Torino, Italy}
\affiliation[d]{Bethe Center for Theoretical Physics, Universit\"at Bonn, D-53115, Germany}
\affiliation[e]{Universit\'e Paris-Saclay, CNRS, IJCLab, 91405 Orsay, France}
\affiliation[f]{Institute for Theoretical Particle Physics, KIT, 76128 Karlsruhe, Germany}
\emailAdd{samuel.abreu@cern.ch}
\emailAdd{matteo.becchetti@unito.it}
\emailAdd{cduhr@uni-bonn.de}
\emailAdd{melih.ozcelik@ijclab.in2p3.fr}

\abstract{We present the analytic expressions for the
 two-loop form factors for the production or decay of
 pseudo-scalar quarkonia, in a scheme where the
 quarks are produced at threshold. We consider the two-loop amplitude for the
 process $\gamma \gamma \leftrightarrow {^1S_0^{[1]}}$, that was previously known
 only numerically, as well as for the processes $gg \leftrightarrow
 {^1S_0^{[1]}}$, $\gamma g \leftrightarrow {^1S_0^{[8]}}$ and
 $gg \leftrightarrow {^1S_0^{[8]}}$, which have not been computed before. The
 two-loop corrections to $gg \leftrightarrow {^1S_0^{[1]}}$ are the last
 missing ingredients for a full NNLO calculation of $\eta_Q$
 hadro-production.  
 We discuss how the singularity structure
 of the amplitudes is affected by the threshold kinematics, 
 which in particular introduces Coulomb singularities. 
 In this context, we first show
 how the usual structure of the infrared singularities degenerates 
 at threshold kinematics, and then extract the anomalous dimensions
 governing the Coulomb singularities for colour-singlet and octet
 channels, the latter being presented here for the first time.
 We give high-precision numerical results for the
 hard functions, which can be used for phenomenological studies of
 $\eta_Q$ production and decay at NNLO.
}

\DeclareMathOperator{\Tr}{Tr}
\newcommand{\beq}{\begin{equation}}

\newcommand{\eeq}{\end{equation}}

\newcommand{\bea}{\begin{eqnarray}}
\newcommand{\eea}{\end{eqnarray}}
\newcommand{\bfig}{\begin{figure}}
\newcommand{\efig}{\end{figure}}
\newcommand{\bc}{\begin{center}}
\newcommand{\ec}{\end{center}}

\newcommand{\msbar}{{$\overline{\text{MS}}$}}

\DeclareMathOperator{\Cl}{Cl}
\DeclareMathOperator{\Li}{Li}
\DeclareMathOperator{\Ree}{Re}
\DeclareMathOperator{\Ime}{Im}

\DeclareMathOperator{\hpli}{HPLI}

\date{}

\begin{document}
\maketitle
\flushbottom


\section{Introduction}
\label{sec:intro}

The high-luminosity program of the Large Hadron Collider (LHC), which will
take place during the second part of this decade, will enable us to study the
fundamental interactions among particles at an unprecedented level of
precision and to measure a large number of physical observables at
the percent level. 
A lot of effort has to be put into improving theoretical predictions
to reach this level of precision in order to make the most of the LHC physics
program.

The production and decay of quarkonium bound states play an important
role within the context of this program. Indeed, quarkonium
physics can be used as a probe to study several aspects of
QCD, such as the interplay between the
perturbative and non-perturbative regimes of QCD 
\cite{Kramer:2001hh,Lansberg:2019adr,QuarkoniumWorkingGroup:2004kpm,
Lansberg:2006dh,Andronic:2015wma} or the analysis of the gluon 
Parton Distribution Function (PDF) of the
proton \cite{Halzen:1984rq,Martin:1987ww,Martin:1987vw,Jung:1992uj,Lansberg:2020ejc}.
Specifically, charmonium production can be used to set constraints on the
PDFs at energy scales on the order of the charm quark mass.  
Quarkonium physics also provides a way to test the convergence of the 
perturbative expansion in QCD, since the strong coupling $\alpha_s$ is not 
so small at the relevant energy scales
(see for instance refs.~\cite{Lansberg:2020ejc,Ozcelik:2021zqt}).

In this paper we focus on the production and decay of a pseudo-scalar
quarkonium state $\eta_Q$, which is a bound state of a quark-antiquark pair
$Q\overline{Q}$, where the massive quark $Q$ can be either a $c$ or $b$ quark. 
The state-of-the-art for this process are next-to-leading order (NLO) QCD 
corrections~\cite{Hagiwara:1980nv,Kuhn:1992qw,Schuler:1994hy,Petrelli:1997ge}. 
An interesting
feature of NLO corrections to pseudo-scalar quarkonium hadro-production is
the appearance of negative cross sections, whose origin can be traced back to
an over-subtraction of the initial-state collinear divergences inside the
PDFs in the \msbar-scheme~\cite{Lansberg:2020ejc}. While it is possible to
devise a prescription of how to avoid the appearance of negative cross
sections at NLO~\cite{Lansberg:2020ejc,Ozcelik:2019qze}, most likely only a complete
next-to-next-to-leading order (NNLO) computation can provide
reliable phenomenological predictions for this process. The NNLO corrections
require the knowledge of the two-loop contributions for
the production of a quarkonium state, which are currently unavailable 
in the literature.

One of the main goals of this paper is to close this gap and to present for
the first time the two-loop QCD corrections to the  amplitudes
for both colour-singlet and colour-octet configurations, in the channels
$\gamma\gamma$, $\gamma g$ and $gg$. More precisely, we will consider the 
processes $\gamma \gamma \leftrightarrow {^1S_0^{[1]}}$, 
$gg \leftrightarrow {^1S_0^{[1]}}$, $\gamma g \leftrightarrow {^1S_0^{[8]}}$
and $gg \leftrightarrow {^1S_0^{[8]}}$. The computation is carried out within
the framework of Non-Relativistic QCD (NRQCD) \cite{Bodwin:1994jh}, where the
production mechanism of the quarkonium state assumes the factorisation into a
perturbative part, which describes the high-energy physics of the process, and
a non-perturbative part, which takes into account the low-energy physics.
While the two-loop corrections to the decay of the colour-singlet state into
two photons have already been calculated numerically \cite
{Czarnecki:2001zc, Feng:2015uha}, the corrections to the other three
processes have not been calculated before and are presented here for the first time.\footnote{The hadronic decay width of $\eta_Q$ has been computed up to 
NNLO in pure numerical form in ref.~\cite{Feng:2017hlu}
using the optical theorem and transforming phase-space
integrations into loop integrals. The results for the two-loop virtual
contributions have, however, not been given, so the results presented in that reference
cannot be used for NNLO pseudo-scalar
quarkonium hadro-production.} Moreover, the two-loop QCD corrections to
the colour-singlet configuration in the $gg$ channel are the last missing
ingredients for a full NNLO computation for pseudo-scalar quarkonium
hadro-production.

The computation of these processes is performed by decomposing the amplitudes
into form factors. Using \emph{Integration-By-Parts (IBP)} identities~\cite{Chetyrkin:1979bj,Chetyrkin:1981qh}, the form factors can be written in terms of a basis of scalar Feynman
integrals, the so-called \emph{master integrals}. The evaluation of the
set of master integrals required for these amplitudes was discussed in
ref.~\cite{Abreu:2022vei}, where we provided both analytic results
and high-precision numerical evaluations. Here we simply note that these
integrals involve multiple polylogarithms (MPLs) \cite{GoncharovMixedTate}
but also elliptic multiple polylogarithms (eMPLs) 
\cite{brown2013multiple,Broedel:2014vla,Broedel:2017kkb} (and the related
iterated integrals of Eisenstein series \cite{ManinModular,Brown:mmv}).
While MPLs are well understood and their analytic manipulation and numerical
evaluation is under good control, the same is not true for their elliptic
generalisation. In particular, the high-precision numerical evaluations
of the integrals involving elliptic functions are not obtained
from their analytic representations, but rather by numerically
solving the differential equations they satisfy with tools such as
$\tt{AMFlow}$ \cite{Liu:2017jxz,Liu:2021wks,Liu:2022chg} 
and $\tt{diffexp}$ \cite{Hidding:2020ytt}.

The paper is structured as follows. In section \ref{sec:setup} we present the
general setup of the computation and we discuss the decomposition of the
amplitudes in terms of form factors. Section \ref{bareAndUV} is dedicated
to the description of the general structure of the bare form factors and
the UV renormalisation procedure. 
In section \ref{sec:IR} we analyse the IR pole structure, including
the Coulomb singularities.
Finally, in section \ref{sec:formfac} we present our results for the finite 
remainder of the form factors for the different processes. 
Our conclusions and outlook are given in section \ref{sec:conclusions}.


\section{Computational setup}
\label{sec:setup}

Within the framework of NRQCD~\cite{Bodwin:1994jh}, the
production of a quarkonium state can be factorised into a perturbative part
that describes the production of a heavy-quark pair $Q\overline{Q}$ at a hard
scale $\mu\sim m_Q$, and a non-perturbative part that describes the
hadronisation of the $Q\overline{Q}$ pair to the bound state $\mathcal{Q}$ at
a much lower scale $\mu_{\Lambda}<m_Q$. This factorisation can be
expressed at the partonic level as
\begin{equation} 
    d\hat{\sigma}_{ab}{\left(\mathcal{Q}+\{k\}\right)}=\sum_n d\hat{\sigma}_{ab}{\left(Q\overline{Q}{\left[n\right]}+
    \{k\}\right)}\langle\mathcal{O}^n_{\mathcal{Q}} \rangle\,,
    \label{eq:factorization}
\end{equation} 
where $a$ and $b$ are the initial-state particles, and
$d\hat{\sigma}_{ab}{\left(Q\overline{Q}[n]+\{k\}\right)}$ describes the
short-distance production of a $Q\overline{Q}$ pair in a given quantum
configuration $n$ with additional partons in the final state represented by
$\{k\}$. The quantum configuration $n$ of the $Q\overline{Q}$ state can be
expressed in spectroscopic notation as ${^{2S + 1} L_J^{[1,8]}}$, where $S$
is the total spin of the $Q\overline{Q}$ pair, $L$ is the orbital angular
momentum and $J$ is the total angular momentum. The superscript
 $[1,8]$ indicates that the $Q\overline{Q}$ pair  is in either a colour-singlet or
colour-octet state. The hadronisation of the
$Q\overline{Q}[n]$ state into the quarkonium state $\mathcal{Q}$ is encoded
in the non-perturbative Long-Distance Matrix Element (LDME) 
$\langle\mathcal{O}^n_{\mathcal{Q}} \rangle$.

While the sum in the factorisation formula eq.~\eqref{eq:factorization}
proceeds over all quantum configurations $n$, in this paper only a few 
contributions will be relevant. Indeed, the factorisation formula admits an
expansion in both the strong coupling $\alpha_s$ and the relative velocity
$v$ between the $Q\overline{Q}$ pair in the rest frame of the quarkonium. We consider only pseudo-scalar $S$-wave states in both colour-singlet and colour-octet configurations, ${^1 S_0^{[1,8]}}$.
More specifically, the colour-singlet state 
${^1S_0^{[1]}}$ is the leading term in the $v$-expansion of $\eta_Q$
production and corresponds to the colour-singlet
model~\cite{Chang:1979nn,Berger:1980ni,Baier:1981uk}. We will also
consider the final-state $Q\overline{Q}$ pair to be in the colour-octet state
${^1S_0^{[8]}}$. For the (short-distance) perturbative corrections, we will always
work at leading order in $v$, that is, we set $v=0$ at the integrand level.

Since we are primarily interested in $\eta_Q$ production, we will briefly
discuss the LDME and its dominant contribution in the ${^1S_0^{[1]}}$ channel.
It can be expressed in terms of the total wave function $\psi_0$ at the
origin \cite{Bodwin:1994jh, Petrelli:1997ge},
\begin{equation}
    \langle \mathcal{O}^{{^1S_0^{[1]}}}_
    {\eta_Q} \rangle=\left\vert \psi_0 \right\vert^2=
    \frac{\left\vert R_0 \right\vert^2}{4\pi},
\end{equation}
where we have also given the relation to the more commonly used
radial wave function at the origin $R_0$ and the
spherical harmonic $Y_{00}=1/\sqrt{4\pi}$. Due to heavy-quark
spin symmetry, the radial wave function $R_0$ is the same for both $\eta_Q$
and $J/\psi$ up to higher-order corrections in the $v$-expansion. $R_0$ can
be computed via the Schr\"odinger equation, and its value can also
be extracted from the leptonic decay width of the
$J/\psi$ \cite{Schuler:1994hy,Kramer:1995nb}.\footnote{There are different models that yield different numerical
values for the radial wave function. For instance, in 
refs.~\cite{Brodsky:2009cf,Lansberg:2020ejc}, the numerical values used for the $S$-wave
functions were $\left\vert R_0\right\vert^2_{\eta_c}=1 \text{ GeV}^3$ and
$\left\vert R_0\right\vert^2_{\eta_b}=7.5 \text{ GeV}^3$.}

The main focus of this paper are the perturbative corrections to 
eq.~\eqref{eq:factorization}, described by
the short-distance interaction 
\begin{equation}\label{eq:hardInt}
a{\left(k_1\right)}b{\left(k_2\right)}\rightarrow Q{\left(p_1\right)}\overline{Q}{\left(p_2\right)}\,,
\end{equation} 
where in our case $a$ and $b$ represent either gluons or photons. 
For the ${^1S_0}^{[1,8]}$ state, we consider the final-state heavy quarks at threshold
kinematics. This corresponds to
\begin{equation}\label{eq:kin1}
k_1^2=k_2^2=0,\quad\quad p^2=\frac{1}{2}k_1\cdot k_2=m_Q^2\,, \quad\text{ with }\quad 
p=p_1=p_2=\frac{1}{2}\left(k_1+k_2\right),
\end{equation}
where the Mandelstam variables are given by
\begin{equation}\label{eq:kin2}
\hat{s}=\left(k_1+k_2\right)^2=M_{\mathcal{Q}}^2=4m_Q^2\,,\quad\quad 
\hat{t}=\left(k_1-p\right)^2=-m_Q^2\,,\quad\quad \hat{u}=\left(k_2-p\right)^2=-m_Q^2\,.
\end{equation} 
This effectively reduces the kinematics underlying the process in \cref{eq:hardInt}
to those of a three-point process.

In this paper we are only interested in the two-loop contributions
to the production or decay of a quarkonium bound state. Specifically, we consider
 the two-loop amplitudes for the channels 
$\gamma \gamma \leftrightarrow {^1S_0^{[1]}}$,
$gg\leftrightarrow {^1S_0^{[1]}}$,
$\gamma g \leftrightarrow {^1S_0^{[8]}}$ and $g g \leftrightarrow{^1S_0^{[8]}}$,
where the double-arrows indicate that we consider both production and decay.
Indeed, the channels with a light quark pair in the initial/final state, $q\overline{q}\leftrightarrow {^1S_0}^{[1,8]}$, are loop-induced and only contribute at NNLO as the product of
one-loop amplitudes. This contribution vanishes in $d=4$ dimensions. 

To compute the required amplitudes, we first
generate the Feynman diagrams with a $Q\overline{Q}$ pair in the final state
using the $\tt{FeynArts}$ package~\cite{Hahn:2000kx}.
We then need to project the $Q\overline{Q}$ pair onto the $^1S_0^{[1,8]}$ state.
All colour and Lorentz algebra manipulations are performed
with $\tt{FeynCalc}$~\cite{Shtabovenko:2016sxi}.
As the amplitude has two fermions in the
final state, it contains the product of spinors
$\overline{u}$, $v$ with matrices $\mathcal{T}$ involving Dirac
$\gamma$ matrices. As this product is a number, we can convert it to a trace
and write
\begin{equation}
\begin{split}
    \overline{u}{\left(p,s_1\right)}\mathcal{T}v{\left(p,s_2\right)}=
    &\Tr{\left[\overline{u}{\left(p,s_1\right)}\mathcal{T}v{\left(p,s_2\right)}\right]}
    =\Tr{\left[\mathcal{T}v{\left(p,s_2\right)}\overline{u}{\left(p,s_1\right)}\right]}.
\end{split}    
\end{equation}
The projection of the final $Q\overline{Q}$ pair onto a pseudo-scalar state
$^1S_0^{[1,8]}$ can be done by means of the replacement 
\cite{Kuhn:1979bb, Guberina:1980dc, Baier:1983va, Gastmans:1987be,
Petrelli:1997ge},\footnote{The relative normalisation of the
LDME and the short-distance part can be chosen freely. 
We follow the conventions of ref.~\cite{Petrelli:1997ge}.}
\begin{equation}
    v{\left(p,s_2\right)}\overline{u}{\left(p,s_1\right)}\rightarrow -\frac{1}{\sqrt{2 m_Q}}\gamma_5 \left(\slashed{p}+m_Q\right) P^{[1,8]}_{ij}.
\end{equation}
The colour-projection operators are
\begin{equation}
    P^{[1,8]}_{ij}=
    \begin{cases} \delta_{ij}/\sqrt{N_c} & \text{colour-singlet }[1]\,, 
    \\ \sqrt{2}\, t_{ij}^a & \text{colour-octet }[8]\,,
    \end{cases}
\end{equation}
where we denote by $t^a$ the generators of the fundamental representation of the $SU(N_c)$ gauge group, and $\delta_{ij}$ is the Kronecker delta. 
The traces involve a $\gamma_5$ matrix, which requires 
a careful treatment when working with dimensional regularisation. 
We employ the 't~Hooft-Veltman scheme \cite{tHooft:1972tcz}.\footnote{However, we observe that, since there is only a single $\gamma_5$ in
the trace, there is no difference when employing naive dimensional
regularisation versus the 't Hooft-Veltman scheme.}

The Lorentz structure of the amplitude for the production of a pseudo-scalar
state is independent of the channel and can be written as\footnote{
The polarisation vectors of the gluons and photons must be
complex conjugated in the amplitudes depending on whether they correspond to the
production or decay channels.}
\begin{equation}
\begin{split}
    \mathcal{A}_{p, c}=&\mathcal{A}_{p, c; \mu \nu}\, \varepsilon^{\mu}{\left(k_1\right)} \varepsilon^{\nu}{\left(k_2\right)}
    =\tilde{\mathcal{A}}_{p, c}\, \epsilon_{\mu \nu \rho \sigma}\, \varepsilon^{\mu}{\left(k_1\right)} \varepsilon^{\nu}{\left(k_2\right)} k_1^{\rho} k_2^{\sigma}.
\end{split}
\end{equation}
where $p$ indicates the channel ($p = gg$, $\gamma g$, $\gamma \gamma$), 
and $c$ denotes the colour state  ($c = [1],[8]$). 
The scalar form factor $\tilde{\mathcal{A}}_{p, c}$ is obtained
with the projection operator
\begin{equation}
    \mathcal{P}^{\mu \nu}=\frac{1}{4\left(d-3\right)\left(d-2\right)m_Q^4} \epsilon^{\mu \nu \rho' \sigma'}\, k_{1, \rho'} k_{2,\sigma'},
\end{equation}
where the overall normalisation is fixed by requiring that
\begin{equation}
\mathcal{P}^{\mu \nu} \mathcal{A}_{p,c; \mu \nu}= \tilde{\mathcal{A}}_{p,c}\,.
\end{equation}

The (bare) scalar form factor $\tilde{\mathcal{A}}_{p, c}$ can be expanded
into powers of the (bare) strong coupling $\alpha^B_s$. We define
the normalised bare form factor $\mathcal{F}_{p, c}$ and its
perturbative expansion as
\begin{equation}
\begin{split}
    \mathcal{F}_{p, c}=&\mathcal{A}_{p, c}/\mathcal{A}_{p, c}^{(0)}
    =\left(\frac{\alpha^B_s}{\pi}\right)^q \left[1 
    + \left(\frac{\alpha^B_s}{\pi}\right) \mathcal{F}_{p, c}^{(1)} 
    + \left(\frac{\alpha^B_s}{\pi}\right)^{2} \mathcal{F}_{p, c}^{(2)}
    +\mathcal{O}{\left(\alpha^B_s\right)^{3}}\right]\,,
\end{split}
\end{equation}
where $\mathcal{A}_{p, c}^{(0)}$ is given by
\begin{equation}
    \mathcal{A}_{p,c}^{(0)}= i \frac{4\pi^2 \sqrt{2}}{m_Q^{5/2}} 
    \mathcal{C}_{p, c}^{\text{col.}}\; 
    \mathcal{C}_{p, c}^{\text{coup.}} \, \epsilon_{\mu \nu \rho \sigma}\, 
    \varepsilon^{\mu}{\left(k_1\right)} \varepsilon^{\nu}{\left(k_2\right)} 
    k_1^{\rho} k_2^{\sigma}\,,
    \label{eq:atildepc}
\end{equation}
and $q=0$ for $ \mathcal{A}_{\gamma\gamma,[1]}$, $q=\frac{1}{2}$ for $ \mathcal
{A}_{\gamma g,[8]}$, and $q=1$ for $ \mathcal{A}_{gg,[1]}$ and 
$ \mathcal{A}_{gg,[8]}$.
The channel-dependent factors $\mathcal{C}_{p, c}^{\text{col.}}$ and 
$\mathcal{C}_{p, c}^{\text{coup.}}$  are given by 
\begin{equation}
    \mathcal{C}_{p, c}^{\text{col.}}=\begin{cases}
    \sqrt{N_c} & \hspace{0.5cm} \gamma \gamma \leftrightarrow {^1S_0^{[1]}}, \\
    T_F\, \delta^{ab}/\sqrt{N_c} & \hspace{0.5cm} gg \leftrightarrow {^1S_0^{[1]}}, \\
    \sqrt{2}\,T_F\, \delta^{bc} & \hspace{0.5cm} \gamma g \leftrightarrow {^1S_0^{[8]}}, \\
    \sqrt{2}\,T_F\, d^{abc}/2 & \hspace{0.5cm} g g \leftrightarrow {^1S_0^{[8]}}, \\
    \end{cases}
    \hspace{0.5cm}
    \mathcal{C}_{p, c}^{\text{coup.}}=\begin{cases}
    e_Q^2 \alpha_{em}/\pi & \hspace{0.5cm} \gamma \gamma \leftrightarrow {^1S_0^{[1]}}, \\
    1 & \hspace{0.5cm} g g \leftrightarrow {^1S_0^{[1]}}, \\
    e_Q \sqrt{\alpha_{em}/\pi} & \hspace{0.5cm} \gamma g 
    \leftrightarrow {^1S_0^{[8]}}, \\
    1 & \hspace{0.5cm} g g \leftrightarrow {^1S_0^{[8]}}, \\
    \end{cases}
\end{equation}
where $e_Q$ denotes the electric charge of the heavy quark and we defined the usual quantities
\begin{equation}
    \begin{split}
    \operatorname{Tr}[t^at^b] =& T_F \delta^{ab}, \hspace{1cm} 
    \operatorname{Tr}[t^at^bt^c] = T_F \frac{1}{2} 
    \left(d^{abc}+i f^{abc}\right)\,.
    \end{split}
\end{equation}
In our conventions, we set $T_F=1/2$. 

The two-loop scalar form factors $\mathcal{F}_{p, c}^{(2)}$ can be decomposed
into a basis of two-loop Feynman integrals. In order to do so, however, we must
first account for partial-fraction relations that arise because of the degenerate
kinematics of \cref{eq:kin1,eq:kin2}. For this we use the package 
$\tt{Apart}$ \cite{Feng:2012iq}. Details and consequences of this 
procedure are given in our companion paper~\cite{Abreu:2022vei}.
Having defined a set of linearly-independent propagators, we
employ standard packages such as $\tt{FIRE}$ \cite{Smirnov:2019qkx} 
or $\tt{KIRA}$ \cite{Klappert:2020nbg} to decompose the form factors into
a basis of 76 master integrals. In ref.~\cite{Abreu:2022vei}, we computed them
both analytically and numerically.

Within this setup, we compute the two-loop form factors 
$\mathcal{F}^{(2)}_{p,c}$ for $\gamma \gamma \leftrightarrow {^1S_0^{[1]}}$, $g
g \leftrightarrow {^1S_0^{[1]}}$, $\gamma g \leftrightarrow {^1S_0^{
[8]}}$ and $g g \leftrightarrow {^1S_0^{[8]}}$. While the first one
had already been computed numerically~\cite{Czarnecki:2001zc, Feng:2015uha},
the last three are obtained here for the first time.
In particular,  $\mathcal{F}_{gg,[1]}^{(2)}$ is the
last missing ingredient for a full NNLO computation of
$\eta_Q$ hadro-production.


\section{The bare amplitude and UV renormalisation}
\label{bareAndUV}

We perform our calculations in the framework of dimensional regularisation, where
the ultraviolet (UV) and infrared (IR) singularities appear as poles in the dimensional
regulator $\epsilon$.
In this section we first discuss the pole structure of the bare form factors
up to two loops, and we then outline the renormalisation procedure
which removes the UV singularities.

\subsection{Bare form factors} \label{sec:bareamp}

One-loop form factors have poles of up to second order 
in the dimensional regulator $\epsilon$. 
We write
\begin{equation}
   \mathcal{F}_{p,c}^{(1)}= S_{\epsilon} \left({m_Q^2}\right)^{-\epsilon}
    \sum_{k \ge -2} \epsilon^{k} \mathcal{F}_{p,c}^{(1,k)}\,,
\end{equation}
where $S_{\epsilon}=\left(4\pi\right)^{\epsilon}e^{-\epsilon \gamma_{E}}$.
With this choice of normalisation, the $m_Q^2$ dependence is fully factorised 
and the coefficients $ \mathcal{F}_{p,c}^{(1,k)}$ are simply numbers.
They can be decomposed in terms of the colour factors as
\begin{equation}
 \mathcal{F}_{p,c}^{(1,k)} = C_A  \mathcal{F}_{p,c;A}^{(1,k)} + C_F  \mathcal{F}_{p,c;F}^{(1,k)}\,,
\end{equation}
where $C_A$ and $C_F$ are the usual Casimir invariants of $SU(N_c)$,
\begin{equation}
t^{a}_{ik} t^a_{kj} = C_F\, \delta_{ij} = \frac{N_c^2-1}{2N_c}\,\delta_{ij}\textrm{~~~and~~~} f^{acd}f^{bcd} = C_A\, \delta^{ab} = N_c\,\delta^{ab}\,.
\end{equation}
The coefficients of the poles in $\epsilon$ are particularly simple.
Indeed, the poles proportional to $C_F$ are identically zero for all form factors
\begin{equation}
\begin{split}
    \mathcal{F}_{p,c;F}^{(1,-2)} =\mathcal{F}_{p,c;F}^{(1,-1)} = 0\,,
    \label{eq:fgamgam_1_F_poles}
\end{split}
\end{equation}
while the poles proportional to $C_A$ are form-factor-dependent and read
\begin{align}
    \mathcal{F}_{\gamma\gamma,[1];A}^{(1,-2)}=&0\,, \quad\quad\quad\,\,   
    \mathcal{F}_{\gamma\gamma,[1];A}^{(1,-1)}=0\,,\label{eq:fgamgam_1_A_poles}
\\
     \mathcal{F}_{gg,[1];A}^{(1,-2)}=&-\frac{1}{2}\,, \quad\quad     
     \mathcal{F}_{gg,[1];A}^{(1,-1)}=-\frac{i\pi}{2}+\log{2}\,,
\\
     \mathcal{F}_{\gamma g,[8];A}^{(1,-2)}=&-\frac{1}{4}\,, \quad\quad    
     \mathcal{F}_{\gamma g,[8];A}^{(1,-1)}=-\frac{1}{4}+\frac{1}{2}\log{2}\,,
\\
    \mathcal{F}_{gg,[8];A}^{(1,-2)}=&-\frac{1}{2}\,, \quad\quad     
    \mathcal{F}_{gg,[8];A}^{(1,-1)}=-\frac{i\pi}{4}-\frac{1}{4}+\log{2}\,.
\end{align}

\begin{figure}
  \begin{center}
    \subfloat[]{\includegraphics[width=4.5cm]{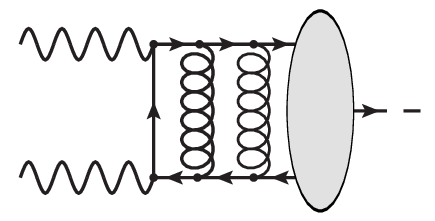}
    \label{fig:gamgamabelian}}\, 
    \subfloat[]{\includegraphics[width=4.9cm]{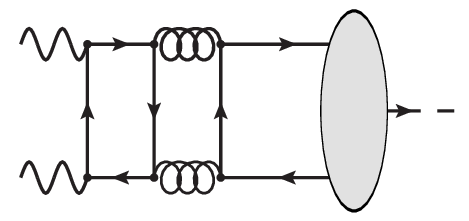} 
    \label{fig:gamgamlbl}}\, 
    \subfloat[]{\includegraphics[width=4.5cm]{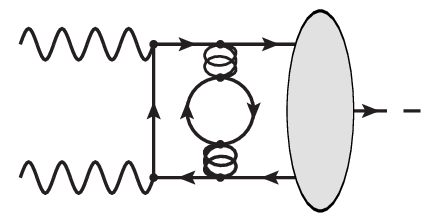} 
    \label{fig:gamgamvac}}
  \end{center}
  \caption{Two-loop diagrams for the form factor 
  $\gamma \gamma \leftrightarrow {^1S_0^{[1]}}$ with 
  (a) regular contributions, (b) light-by-light contributions and 
  (c) vacuum polarisation contributions.}
  \label{fig:typesdiagrams}
\end{figure}

The two-loop form factors have poles up to order $\epsilon^{-4}$.
We write the form factor as
\begin{equation}
   \mathcal{F}_{p,c}^{(2)}= S^2_{\epsilon} \left(m_Q^2\right)^{-2\epsilon}
    \sum_{k \ge -4} \epsilon^{k} \mathcal{F}_{p,c}^{(2,k)}.
\end{equation}
We find it convenient to classify the different contributions that appear in the  two-loop amplitude. First, we distinguish terms that survive in the limit $C_A\to0$, which we call abelian contributions, and terms that vanish. Second, we distinguish sets of \emph{gauge-invariant} contributions: the \emph{regular} two-loop contributions, coming from diagrams without closed fermion loops, the \emph{light-by-light scattering} contributions, coming from diagrams with fermion loops connected to the external bosons, and the \emph{vacuum polarisation} contributions, coming from diagrams with closed fermion loops in gluon propagators and with triple gluon vertices. Representative diagrams for each contribution can be found in \cref{fig:typesdiagrams}. We can express the bare two-loop amplitude as
\begin{equation}
    \mathcal{F}_{p,c}^{(2,k)}=\mathcal{F}_{p,c;\text{reg}}^{(2,k)}+\mathcal{F}_{p,c;\text{lbl}}^{(2,k)}+\mathcal{F}_{p,c;\text{vac}}^{(2,k)},
\end{equation}
where
\begin{align}
    \mathcal{F}_{p,c;\text{reg}}^{(2,k)}=&C_F^2\, \mathcal{F}_{p,c;FF}^{(2,k)} + C_F C_A\, \mathcal{F}_{p,c;FA}^{(2,k)} + C_A^2\, \mathcal{F}_{p,c;AA}^{(2,k)}\,,
    \label{eq:bareregval}
    \\[4pt]
    \begin{split}
    \mathcal{F}_{p,c;\text{lbl}}^{(2,k)}=&C_F T_F n_h\, \mathcal{F}_{p,c;Fh;\text{lbl}}^{(2,k)} + C_F T_F \tilde{n}_l\, \mathcal{F}_{p,c;Fl;\text{lbl}}^{(2,k)}
    \\
    &+ C_A T_F n_h\, \mathcal{F}_{p,c;Ah;\text{lbl}}^{(2,k)} + C_A T_F \tilde{n}_l\, \mathcal{F}_{p,c;Al;\text{lbl}}^{(2,k)}\,,
    \end{split}
    \label{eq:barelblval}
    \\[4pt]
    \begin{split}
    \mathcal{F}_{p,c;\text{vac}}^{(2,k)}=&C_F T_F n_h\, \mathcal{F}_{p,c;Fh;\text{vac}}^{(2,k)} + C_F T_F n_l\, \mathcal{F}_{p,c;Fl;\text{vac}}^{(2,k)}
    \\
    &+ C_A T_F n_h\, \mathcal{F}_{p,c;Ah;\text{vac}}^{(2,k)} + C_A T_F n_l\, \mathcal{F}_{p,c;Al;\text{vac}}^{(2,k)}\,,
    \end{split}
    \label{eq:barevacval}
\end{align}
where $n_h$ and $n_l$ are the number of heavy and light quarks respectively. For the light-by-light contributions, we have to define the quantity $\tilde{n}_l$ that takes into account the QED coupling between the external photons and the fermion flavour inside the loop. This quantity reads
\begin{equation}
    \tilde{n}_l=\begin{cases} \sum_{i}^{n_l} e_i^{2}/e_Q^{2} & \hspace{1cm} \text{for}\;\; \gamma \gamma \;\;\;\text{channel}\,, \\
    \sum_{i}^{n_l} e_i/e_Q & \hspace{1cm} \text{for}\;\; \gamma g \;\;\; \text{channel}\,, \\
    n_l & \hspace{1cm} \text{for}\;\; gg \;\;\; \text{channel}\,.
    \end{cases}
    \label{eq:definitionnltilde}
\end{equation}
We further note that the light-by-light contributions are finite in four dimensions and are thus not affected by the procedure of UV renormalisation. In appendix~\ref{sec:barepole}, we give, in addition to the analytic expressions for the poles, also the numerical values for the finite part for the contributions given in eqs.~\eqref{eq:bareregval}-\eqref{eq:barevacval}. We observe that, while the two-loop form factor $\mathcal{F}^{(2)}_{\gamma\gamma,[1]}$ has poles of at most second order, the other form factors have poles starting at the quadruple pole.

It is clear that abelian contributions should be very similar across different
channels. Indeed, they are only different in the light-by-light contributions, 
where colour-singlet channels differ from colour-octet channels by a factor of 2 coming from the different colour algebra. Because the light-by-light
contributions are finite, we find that the abelian contributions to the pole 
structure of all channels is the same. Verifying that these relations hold
provides a stringent check of our calculations.
We also note that in the limit 
$C_A\rightarrow 0$, $C_F\rightarrow 1$ and $T_F\rightarrow 1$ the 
colour-singlet contributions should reproduce the two-loop contributions
to para-positronium production or decay obtained numerically in
ref.~\cite{Czarnecki:1999ci}, which provides another important check.


\subsection{UV renormalisation}  \label{sec:UV}

Having explained how we obtained the bare two-loop form factors $\mathcal{F}_{p,c}$, we 
now discuss how to compute their renormalised counterparts. 
We work in the on-shell renormalisation scheme
for the heavy-quark wave function, for the heavy-quark mass and for the gluon
wave function. As for the strong coupling $\alpha_s$, we employ
the \msbar-scheme. The renormalisation is performed with multiplicative
factors $Z_\kappa$, with $\kappa=Q$, $m$, $g$, $\alpha_s$ respectively. For instance,
the bare coupling is related to the renormalised coupling $\alpha_s$ by
\begin{equation}\label{eq:aRun}
    \alpha_s^B=S_{\epsilon}^{-1}\mu_R^{2\epsilon}Z_{\alpha_s}\alpha_s^{\left(n_f\right)},
\end{equation}
where we take into account $n_f=n_l+n_h$ flavours in the running of the coupling.
The $Z_\kappa$ factors admit an expansion in the renormalised coupling with $n_f$ flavours as
\begin{equation}
    Z_\kappa=1+\left(\frac{\alpha_s^{\left(n_f\right)}}{\pi}\right)Z_\kappa^{(1)}
    +\left(\frac{\alpha_s^{\left(n_f\right)}}{\pi}\right)^2 Z_\kappa^{(2)}+\mathcal{O}{\left(\alpha_s^3\right)}\,,
    \label{eq:Ziexpansion}
\end{equation}
and the $Z^{(i)}_\kappa$ are collected in appendix~\ref{sec:RenCoef}. 
It is more common to express the results in terms of a coupling
$\alpha_s^{\left(n_l\right)}$ where we only consider the light-quark
flavours in the running of the coupling. In order to convert
from one coupling to the other, we apply the decoupling 
identity~\cite{Bernreuther:1981sg}
\begin{equation}
    \alpha_s^{\left(n_l+n_h\right)}=\zeta_{\alpha_s}\alpha_s^{\left(n_l\right)},
\end{equation}
where $\zeta_{\alpha_s}$ admits an expansion in the strong coupling 
similar to \cref{eq:Ziexpansion}, but with $n_l$ flavours in the running of the coupling. The coefficients for $\zeta_{\alpha_s}$ are also given in appendix~\ref{sec:RenCoef}.

As done for the other bare quantities, the renormalisation of the heavy-quark
mass $m_Q$ could in principle be implemented through a simple replacement
$m_Q^{B}=Z_m\, m_Q$ in the amplitude. However, given the degenerate
kinematics underlying our process (cf. \cref{eq:kin1,eq:kin2}), we have
evaluated the integrals at $\hat{s}=4m_Q^2$. Hence the threshold value of
$\hat{s}$ is related to the on-shell mass $m_Q$ of the heavy quarks, while the propagators involve the bare mass $m_Q^B$.
Since we did not distinguish between the bare and on-shell masses at the time of the diagram generation, it is not possible to simply substitute
$m_Q^B$ by its renormalised value. Instead, we compute counterterms that are added to the bare amplitude to implement the heavy-quark mass
renormalisation. This involves computing one-loop amplitudes with doubled
propagators, which we do using the same standard approach described above
for the calculation of the bare amplitudes.

We write the renormalised form factors, expanded in powers
of $\alpha_s^{\left(n_l\right)}$, as
\begin{equation}
    \mathcal{\overline{F}}_{p, c}=\left(\frac{\alpha^{(n_l)}_s}{\pi}\right)^q \left[1 
    + \left(\frac{\alpha^{(n_l)}_s}{\pi}\right) \mathcal{\overline{F}}_{p, c}^{(1)} 
    + \left(\frac{\alpha^{(n_l)}_s}{\pi}\right)^{2} \mathcal{\overline{F}}_{p, c}^{(2)}
    +\mathcal{O}{\left(\alpha^{(n_l)}_s\right)^{3}}\right],
\end{equation}
where the $n$-loop renormalised form factors can be written as
\begin{equation}
   \mathcal{\overline{F}}_{p,c}^{(n)}=\mu_R^{2n\epsilon}S_{\epsilon}^{-n}
   \mathcal{F}_{p,c}^{(n)}+\mathcal{F}_{p,c}^{(n,\text{CT})}+
   \mathcal{F}_{p,c}^{(n,\text{decoupling})}\,.
   \label{eq:uvrenormalisedformfactor}
\end{equation}
The contribution of all renormalisation factors is collected in 
$\mathcal{F}_{p,c}^{(n,\text{CT})}$. 
Both $\mathcal{F}_{p,c}^{(n)}$ and $\mathcal{F}_{p,c}^{(n,\text{CT})}$
are computed as an expansion in $\alpha_s^{\left(n_f\right)}$
and $\mathcal{F}_{p,c}^{(n,\text{decoupling})}$ translates the result
to an expansion in $\alpha^{(n_l)}_s$.

At one-loop level, the counterterm contribution in 
\cref{eq:uvrenormalisedformfactor} gives 
\begin{equation}
   \mathcal{F}_{p,c}^{(1,\text{CT})}= 
   q\left(Z_g^{(1)}+Z_{\alpha_s}^{(1)}\right)+Z_Q^{(1)}-Z_m^{(1)}\,.
\end{equation}
We note that $Z_Q^{(1)}=Z_m^{(1)}$, and the renormalised form factor 
$\mathcal{\overline{F}}^{(1)}_{\gamma\gamma,[1]}$ 
(for which $q=0$) equals its bare counterpart.
As will be discussed below, this form factor exhibits neither soft 
nor collinear singularities and is thus finite, which
agrees with our results, see eqs.~\eqref{eq:fgamgam_1_F_poles} and 
\eqref{eq:fgamgam_1_A_poles}.
At two-loop level, the counterterm contribution in \cref{eq:uvrenormalisedformfactor} reads
\begin{equation}
\begin{split}
     \mathcal{F}_{p,c}^{(2,\text{CT})}&=S_{\epsilon}^{-1} \mu_R^{2\epsilon} 
     \mathcal{F}_{p,c}^{(1)} \left[q Z_g^{(1)}+\left(1+q\right)Z_{\alpha_s}^{(1)}
     +Z_Q^{(1)}\right]-Z_m^{(1)}\mathcal{F}_{p,c}^{(1,\text{mass CT})}
    \\
    &+q Z_{\alpha_s}^{(1)}\left(q Z_g^{(1)}+Z_Q^{(1)}-Z_m^{(1)}\right)
    +q Z_{\alpha_s}^{(2)}+q Z_g^{(2)}+q Z_g^{(1)}\left(Z_Q^{(1)}-Z_m^{(1)}\right)
    \\
    &+ \frac{1}{2}q\left(q-1\right)\left[\left(Z_{\alpha_s}^{(1)}\right)^2
    +\left(Z_{g}^{(1)}\right)^2\right]+Z_Q^{(2)}-Z_m^{(2)}-Z_m^{(1)} Z_Q^{(1)}
    +\frac{1}{2}\left(Z_m^{(1)}\right)^2\,,
\end{split}
\end{equation}
where $\mathcal{F}_{p,c}^{(1,\text{mass CT})}$ is obtained by considering
the one-loop amplitude with all possible ways of squaring the massive-quark propagator
(see, e.g., refs.~\cite{Bonciani:2010mn,Czarnecki:1999ci}). These must be computed to
$\mathcal{O}{\left(\epsilon\right)}$ because  $Z_m^{(1)}$ has a simple pole in $\epsilon$.
As for the decoupling contribution in \cref{eq:uvrenormalisedformfactor}, we have
\begin{align}
    \mathcal{F}_{p,c}^{(1,\text{decoupling})}=&\,
    q \, \zeta_{\alpha_s}^{(1)},
    \\
    \mathcal{F}_{p,c}^{(2,\text{decoupling})}=&\,
    q \,\zeta_{\alpha_s}^{(2)}+\frac{1}{2}q
    \left(q-1\right)\left(\zeta_{\alpha_s}^{(1)}\right)^2+
    \left(q+1\right)\zeta_{\alpha_s}^{(1)}\left(\mathcal{\overline{F}}_{p,c}^{(1)}-
    \mathcal{F}_{p,c}^{(1,\text{decoupling})}\right).
\end{align}

The renormalised form factors $\overline{\mathcal{F}}^{(n)}_{p,c}$ are free of UV singularities, 
but still exhibit IR singularities in $\epsilon$, which will be discussed in the next section.
We write the renormalised one-loop form factors as
\begin{equation}
\mathcal{\overline{F}}_{p,c}^{(1)}=  \; \sum_{k \ge -2} \epsilon^{k} \mathcal{\overline{F}}_{p,c}^{(1,k)},
\end{equation}
where we have expanded out all factors depending on $\epsilon$.  
As the renormalisation procedure introduces a $T_F n_l$ term, the colour decomposition now involves
\begin{equation}
\mathcal{\overline{F}}_{p,c}^{(1,k)}=C_A\, 
\mathcal{\overline{F}}_{p,c;A}^{(1,k)}
+C_F\,\mathcal{\overline{F}}_{p,c;F}^{(1,k)}
+T_F n_l\, \mathcal{\overline{F}}_{p,c;l}^{(1,k)}.
\end{equation}
As for the bare amplitudes, in all channels, there are no poles proportional to $C_F$.
For the $C_A$ contributions, we have that
\begin{align}
\begin{split}
    \mathcal{\overline{F}}_{\gamma\gamma,[1];A}^{(1,-2)} =&0\,, 
    \quad\quad\quad\,\, \mathcal{\overline{F}}_{\gamma\gamma,[1];A}^{(1,-1)} =0\,,\\
    \mathcal{\overline{F}}_{gg,[1];A}^{(1,-2)}=&-\frac{1}{2}\,, \quad\quad    
    \mathcal{\overline{F}}_{gg,[1];A}^{(1,-1)}=-\frac{i\pi}{2}+\log{2}-\frac{11}{12}-\frac{1}{2}l_{\mu_R}\,,\\
    \mathcal{\overline{F}}_{\gamma g,[8];A}^{(1,-2)}=&-\frac{1}{4}\,, \quad\quad   
    \mathcal{\overline{F}}_{\gamma g,[8];A}^{(1,-1)}=-\frac{17}{24}+\frac{1}{2}\log{2}-\frac{1}{4}l_{\mu_R}\,,
    \\
    \mathcal{\overline{F}}_{gg,[8];A}^{(1,-2)}=&-\frac{1}{2}\,, \quad\quad    
    \mathcal{\overline{F}}_{gg,[8];A}^{(1,-1)}=-\frac{i\pi}{4}-\frac{7}{6}+\log{2}-\frac{1}{2}l_{\mu_R}\,,
\end{split}
\end{align}
where we used the shorthand notation
\begin{equation}\label{eq:lmu}
l_{\mu}=\log\frac{\mu^2}{m_Q^2}\,.
\end{equation}
Finally, for the $T_F n_l$ contributions the poles are
\begin{align}
\begin{split}
    \mathcal{\overline{F}}_{\gamma\gamma,[1];l}^{(1,-2)}=&0\,, 
    \quad\quad \mathcal{\overline{F}}_{\gamma\gamma,[1];l}^{(1,-1)}=0\,, \\
    \mathcal{\overline{F}}_{gg,[1];l}^{(1,-2)}=&0\,, \quad\quad    
    \mathcal{\overline{F}}_{gg,[1];l}^{(1,-1)}=\frac{1}{3}\,, \\
    \mathcal{\overline{F}}_{\gamma g,[8];l}^{(1,-2)}=&0\,, \quad\quad   
    \mathcal{\overline{F}}_{\gamma g,[8];l}^{(1,-1)}=\frac{1}{6}\,, \\
    \mathcal{\overline{F}}_{gg,[8];l}^{(1,-2)}=&0\,, \quad\quad    
    \mathcal{\overline{F}}_{gg,[8];l}^{(1,-1)}=\frac{1}{3}\,.
\end{split}
\end{align}

The two-loop form factors can be similarly written as
\begin{equation}
\mathcal{\overline{F}}_{p,c}^{(2)}=  \; \sum_{k \ge -4} \epsilon^{k} \mathcal{\overline{F}}_{p,c}^{(2,k)},
\end{equation}
where the $\mathcal{\overline{F}}_{p,c}^{(2,k)}$ can be decomposed into the different sets as
\begin{equation}
    \mathcal{\overline{F}}_{p,c}^{(2,k)}=\mathcal{\overline{F}}_{p,c;\text{reg}}^{(2,k)}+\mathcal{\overline{F}}_{p,c;\text{lbl}}^{(2,k)}+\mathcal{\overline{F}}_{p,c;\text{vac}}^{(2,k)},
\end{equation}
where
\begin{align}
    \mathcal{\overline{F}}_{p,c;\text{reg}}^{(2,k)}=&C_F^2\, \mathcal{\overline{F}}_{p,c;FF}^{(2,k)} + C_F C_A\, \mathcal{\overline{F}}_{p,c;FA}^{(2,k)} + C_A^2\, \mathcal{\overline{F}}_{p,c;AA}^{(2,k)}\,,
    \\[4pt]
    \begin{split}
    \mathcal{\overline{F}}_{p,c;\text{lbl}}^{(2,k)}=&C_F T_F n_h\, \mathcal{\overline{F}}_{p,c;Fh;\text{lbl}}^{(2,k)} + C_F T_F \tilde{n}_l\, \mathcal{\overline{F}}_{p,c;Fl;\text{lbl}}^{(2,k)}
    \\
    &+ C_A T_F n_h\, \mathcal{\overline{F}}_{p,c;Ah;\text{lbl}}^{(2,k)} + C_A T_F \tilde{n}_l\, \mathcal{\overline{F}}_{p,c;Al;\text{lbl}}^{(2,k)}\,,
    \end{split}
    \\[4pt]
    \begin{split}
    \mathcal{\overline{F}}_{p,c;\text{vac}}^{(2,k)}=&C_F T_F n_h\, \mathcal{\overline{F}}_{p,c;Fh;\text{vac}}^{(2,k)} + C_F T_F n_l\, \mathcal{\overline{F}}_{p,c;Fl;\text{vac}}^{(2,k)}
    \\
    &+ C_A T_F n_h\, \mathcal{\overline{F}}_{p,c;Ah;\text{vac}}^{(2,k)} + C_A T_F n_l\, \mathcal{\overline{F}}_{p,c;Al;\text{vac}}^{(2,k)} + T_F^2 n_l^2\, \mathcal{\overline{F}}_{p,c;ll}^{(2,k)}\,.
    \end{split}
\end{align}
The last term proportional to $T_F^2$ is a new colour structure that arises through the renormalisation factors. Since the light-by-light contributions are finite, we have that $\mathcal{\overline{F}}_{p,c;\text{lbl}}^{(2,0)}=\mathcal{F}_{p,c;\text{lbl}}^{(2,0)}$.
As done for the bare form factors, we have collected the singular parts and the finite piece of the renormalised form factors in appendix~\ref{sec:renpole}. For most channels we find that the pole structure is what would be expected for a two-loop amplitude involving external massless particles, that is we find poles up to order $\mathcal{O}{\left(\epsilon^{-4}\right)}$. The exception is the form factor $\mathcal{\overline{F}}_{\gamma\gamma,[1]}$, which has a much simpler pole structure, namely a simple pole with contributions proportional to $C_F^2$ and $C_A C_F$. This pole has a special interpretation that will be discussed in the next section.


\section{Infrared singularities} \label{sec:IR}

The pole structure of renormalised amplitudes in NRQCD
is more involved than that of amplitudes in full QCD.
Indeed, in NRQCD a new type of singularity arises, the so-called
\emph{Coulomb singularity}, see, e.g., 
refs.~\cite{Beneke:1997jm,Czarnecki:1997vz,Czarnecki:2001zc,Kniehl:2006qw,Feng:2015uha}. 
It appears as a consequence of the fact that we have expanded the amplitude 
with respect to the relative velocity $v$ between the heavy quarks. Taking this fact
into account, we define a finite remainder $\mathcal{F}_{p,c}^{\text{fin}}$ as
\begin{equation}
    \mathcal{F}_{p,c}^{\text{fin}}=\mathbf{Z}^{-1}_{\text{Coul.}}
    \mathbf{Z}^{-1}_{\text{IR}}\,\mathcal{\overline{F}}_{p,c}\,,
    \label{eq:defFinRem}
\end{equation} 
where $\mathbf{Z}_{\text{Coul.}}$ is the factor that removes the Coulomb singularity, 
while $\mathbf{Z}_{\text{IR}}$ subtracts the standard infrared (IR) poles. 
They are in general matrices in colour space.

The $\mathbf{Z}_\kappa$ factors above admit an expansion in powers of 
$\alpha^{(n_l)}_s$, similarly to the renormalisation factors discussed in the
previous section. However, while the strong coupling in the renormalisation
factors is evaluated at the renormalisation scale, $\mu_R$, the coupling
expansion in the $\mathbf{Z}_{\text{IR}}$ and $\mathbf{Z}_{\text
{Coul.}}$ factors proceeds at different scales, namely the factorisation
scale, $\mu_F$, and the NRQCD scale, $\mu_{\Lambda}$, respectively. In order
to match the coupling expansions, we will therefore first need to evolve all
couplings to the same scale, for instance the renormalisation scale, $\mu_R$.
Starting from the evolution equation for the strong coupling in
$d=4-2\epsilon$ dimensions,
\begin{equation}
\begin{split}
    \mu^2\frac{\partial \alpha_s}{\partial \mu^2}=&\beta{\left(\alpha_s\right)}-\epsilon \alpha_s
    =-\alpha_s\left[\sum_{n=0}^{\infty} \beta_n \left(\frac{\alpha_s}{4\pi}\right)^{n+1}\right]-\epsilon \alpha_s\,,
\end{split}
\end{equation}
and using the short-hand notation 
$\tilde{\alpha}_s=\alpha_s{\left(\tilde{\mu}\right)}$ 
and $\alpha_s=\alpha_s{\left(\mu\right)}$, we can evolve the 
coupling from the scale $\tilde{\mu}$ to the scale $\mu$ in $d=4-2\epsilon$ 
dimensions with
\begin{equation}
    \tilde{\alpha}_s= \alpha_s \left(\frac{\mu^2}{\tilde{\mu}^2}\right)^{\epsilon}\left[1+\frac{\alpha_s}{\pi} \frac{\beta_0}{4\epsilon} \left(\left(\frac{\mu^2}{\tilde{\mu}^2}\right)^{\epsilon}-1\right)\right]+\mathcal{O}{\left(\alpha_s^3\right)}.
    \label{eq:alphasevolution}
\end{equation}
Expanding \cref{eq:defFinRem} in powers of $\alpha^{(n_l)}_s$ and using the scale 
evolution of the couplings in \cref{eq:alphasevolution}, we then find that
\begin{equation}
\begin{split}\label{eq:genIR}
    \mathcal{F}_{p,c}^{\text{fin}}=&\left(\frac{\alpha_s^{\left(n_l\right)}}{\pi}\right)^{q}
    \left\lbrace 1+\left(\frac{\alpha_s^{\left(n_l\right)}}{\pi}\right)
    \left[\mathcal{\overline{F}}_{p,c}^{(1)}-\left(\frac{\mu_R^2}{\mu_F^2}\right)^{\epsilon}\mathbf{Z}^{(1)}_{\text{IR}}\right] \right.
    \\
    & +\left(\frac{\alpha_s^{\left(n_l\right)}}{\pi}\right)^{2}
    \Bigg[\mathcal{\overline{F}}_{p,c}^{(2)}-\left(\frac{\mu_R^2}{\mu_F^2}\right)^{\epsilon} \mathbf{Z}^{(1)}_{\text{IR}}\left(\mathcal{\overline{F}}_{p,c}^{(1)}+\frac{\beta_0}{4\epsilon} \left(\left(\frac{\mu_R^2}{\mu_F^2}\right)^{\epsilon}-1\right)\right)
    \\
    & \hspace{2.0cm} \left. -\left(\frac{\mu_R^2}{\mu_F^2}\right)^{2\epsilon}\left( \mathbf{Z}^{(2)}_{\text{IR}}-\left(\mathbf{Z}^{(1)}_{\text{IR}}\right)^2\right) 
    -\left(\frac{\mu_R^2}{\mu_{\Lambda}^2}\right)^{2\epsilon} \mathbf{Z}^{(2)}_{\text{Coul.}}\Bigg] \right\rbrace
    \\
    &+\mathcal{O}{\left(\alpha_s^{q+3}\right)}\,,
\end{split}
\end{equation} 
where the quantities $\mathbf{Z}^{(i)}_{\kappa}$ correspond to
the coefficients of $\mathbf{Z}_\kappa$ expanded around the coupling at the
respective scales, $\mu_F$ and $\mu_{\Lambda}$. In addition, we used the
fact that $\mathbf{Z}^{(1)}_{\text{Coul.}}=0$ 
\cite{Czarnecki:2001zc, Kniehl:2006qw, Feng:2015uha}.

In this section we will discuss how to determine the 
$\mathbf{Z}^{(i)}_{\kappa}$. While some of the
ingredients were known in the literature, some are obtained here for the
first time. We will generically label all
scales with $\mu$, however, it is implicitly understood that $\mu=\mu_F$
when discussing $\mathbf{Z}_{\text{IR}}$ and $\mu=\mu_{\Lambda}$
when discussing $\mathbf{Z}_{\text{Coul.}}$.

\subsection{General structure of IR singularities} \label{sec:CanIR}

Let us first focus on $\mathbf{Z}_{\text{IR}}$, which describes
the infrared structure of loop amplitudes in QCD \cite{Catani:1998bh,Becher:2009cu,Becher:2009kw,Becher:2009qa,Ferroglia:2009ii}.
It satisfies the evolution equation
\begin{equation}
\frac{d}{d\log{\mu}}\mathbf{Z}_{\text{IR}}=-\mathbf{\Gamma}\,\mathbf{Z}_{\text{IR}}\,,
\label{eq:difZIR}
\end{equation}
where the soft anomalous dimension $\mathbf{\Gamma}$ is a matrix in colour space which admits the perturbative expansion
\begin{equation}
    \mathbf{\Gamma}=\sum_{k=0}^\infty\mathbf{\Gamma}_k \left(\frac{\alpha_s^{\left(n_l\right)}}{\pi}\right)^{k+1}.
    \label{eq:gammaIRexpansion}
\end{equation}
Solving eq.~\eqref{eq:difZIR} order by order in $\alpha_s^{(n_l)}$, we can express $\mathbf{Z}_{\text{IR}}$ as
\begin{equation}
\begin{split}
    \mathbf{Z}_{\text{IR}}=& 1+\left(\frac{\alpha_s^{\left(n_l\right)}}{\pi}\right)
    \left[\frac{\mathbf{\Gamma}'_0}{4\epsilon^2}+\frac{\mathbf{\Gamma}_0}{2\epsilon}\right]
    \\
    &+\left(\frac{\alpha_s^{\left(n_l\right)}}{\pi}\right)^2 
    \left[\frac{\left(\mathbf{\Gamma}'_0\right)^2}{32\epsilon^4}+
    \frac{\mathbf{\Gamma}'_0}{8\epsilon^3}\left(\mathbf{\Gamma}_0-\frac{3}{8}\beta_0\right)+
    \frac{\mathbf{\Gamma}_0}{8\epsilon^2}\left(\mathbf{\Gamma}_0-\frac{1}{2}\beta_0\right)+
    \frac{\mathbf{\Gamma}'_1}{16\epsilon^2}+\frac{\mathbf{\Gamma}_1}{4\epsilon}\right]
    \\
    &+\mathcal{O}{\left(\alpha_s^3\right)}\,,
\end{split}
\end{equation}
where we defined
\begin{equation}
    \mathbf{\Gamma}'=\frac{\partial}{\partial \log{\mu}}\mathbf{\Gamma}\,.
\end{equation}
The explicit form of the soft anomalous dimension matrix $\mathbf{\Gamma}$, 
and therefore of the operator $\mathbf{Z}_{\text{IR}}$, 
is known up to two-loop order \cite{Ferroglia:2009ii,Becher:2009qa,Becher:2009kw,Becher:2009cu}:
\begin{equation}
\begin{split}
    \mathbf{\Gamma}=&\sum_{\left(i,j\right)} \frac{\mathbf{T}_i \cdot \mathbf{T}_j}{2} 
    \gamma_{\text{cusp}} \log{\left(\frac{\mu^2}{-s_{ij}}\right)}+\sum_{i} \gamma^i
    \\
    &-\sum_{\left(I, J\right)} \frac{\mathbf{T}_I \cdot \mathbf{T}_J}{2} 
    \gamma_{\text{cusp}}{\left(\beta_{IJ}\right)}+\sum_{I} \gamma^I +
    \sum_{I,j} \mathbf{T}_I \cdot \mathbf{T}_j \gamma_{\text{cusp}} 
    \log{\left(\frac{m_{I} \mu}{-s_{I j}}\right)}
    \\
    &+\sum_{\left(I,J,K\right)} i f^{abc}\, \mathbf{T}_I^a \mathbf{T}_J^b 
    \mathbf{T}_K^c F_1{\left(\beta_{I J},\beta_{JK}, \beta_{K I}\right)}
    \\
    &+\sum_{\left(I,J\right)} \sum_{k} i f^{abc}\, \mathbf{T}_I^a \mathbf{T}_J^b 
    \mathbf{T}_k^c f_2{\left(\beta_{I J},
    \log{\left(\frac{-\sigma_{J k} v_J \cdot p_k}{-\sigma_{I k} v_I \cdot p_k}\right)}\right)}
    +\mathcal{O}{\left(\alpha_s^3\right)}\,,
     \label{eq:anodimmatrix}
\end{split}
\end{equation}
and 
\begin{equation}
    \mathbf{\Gamma}'=-\gamma_{\text{cusp}}\sum_i \mathbf{T}_i^2\,.
\end{equation}
The sums in eq.~\eqref{eq:anodimmatrix} run over colourful initial- and final-state partons,
and when summing over several parton indices we take them to be distinct (that is, for instance, $i\neq j$ in the first term of the first line).
The lowercase indices $i$ stand for massless partons and the uppercase indices $I$ for massive ones, which in particular implies that the third line in \cref{eq:anodimmatrix} does not contribute in our case as there are only two massive quark legs. The $\mathbf{T}_i^a$ are the generators of the Lie algebra of the gauge group $SU(N_c)$ in the representation of parton $i$. Specifically, we distinguish three different cases. 
When the parton $i$ is a gluon, we have $\left(\mathbf{T}^a_i\right)_{bc}=-i f^{a b c}$. In the case of an initial-state  quark or final-state anti-quark we have that $\left(\mathbf{T}^a_{i}\right)_{\alpha \beta}=-t^{a}_{\beta \alpha}$. Finally, in the case of the emission of a gluon from an initial-state anti-quark or final-state quark we have that $\left(\mathbf{T}^a_{i}\right)_{\alpha \beta}=t^{a}_{\alpha \beta}$. These relations equally apply to massive partons $I$ as heavy (anti-)quarks. It should be kept in mind that the sum over the different colours in eq.~\eqref{eq:anodimmatrix} is performed implicitly. We have the following properties
\begin{align}
    \mathbf{T}_i\cdot \mathbf{T}_j=&\mathbf{T}_j\cdot \mathbf{T}_i,
    \\
    \begin{split}
    \textbf{T}_i^2 =& \begin{cases} 
    C_A & \text{if }i \text{ is a gluon}, \\
    C_F & \text{if }i \text{ is a quark or anti-quark\,,}
    \end{cases}
    \label{eq:Ti2}
    \end{split}
\end{align}
and similarly for massive partons $I,J$.
The expressions for the quark and gluon anomalous dimensions $\gamma^Q$, $\gamma^g$ and for the massless cusp anomalous dimension
$\gamma_{\text{cusp}}$ are collected in \cref{sec:IRRenCoef}. The kinematical dependence in \cref{eq:anodimmatrix} is encoded in the quantities (the indices $a$ and $b$ can denote either massive or massless partons)
\begin{equation}
    s_{ab}=2\sigma_{a b} p_a p_b+i 0^{+},
\end{equation}
with $\sigma_{ab}=+1$ if both partons $a$ and $b$ are both incoming/outgoing 
and $\sigma_{ab}=-1$ otherwise. The cusp anomalous dimension depends on the angle $\beta_{IJ}$, related to the invariants $s_{IJ}$ by
$\cosh{\beta_{IJ}}=-s_{IJ}/\left(2 m_I m_J\right)$, and $v_I$ is defined as $p_I/m_I$.

Equation~\eqref{eq:anodimmatrix} is the general expression for the soft anomalous dimension for any number of external legs up to two loops in full QCD. In our case we can simplify this expression further. 
First we note that, as already mentioned, the third line does not contribute, and neither does the fourth line. Second, the kinematical variables only depend on the mass of the partons and are the same for all form factors, that is
\begin{align}
    s_{ij}=4m_Q^2 + i 0^{+}\,,\qquad
    s_{IJ}=2m_Q^2 + i 0^{+}\,,\qquad
    s_{Ij}=-2m_Q^2 + i 0^{+}\,,
\end{align}
yielding
\begin{align}
    \log{\left(\frac{\mu^2}{-s_{ij}}\right)}=l_{\mu}-2\log{2}+i\pi,\,\qquad
    \log{\left(\frac{m_I \mu}{-s_{Ij}}\right)}=\frac{1}{2}l_{\mu}-\log{2}\,.
\end{align}
We can then write
\begin{equation}
\begin{split}
    \mathbf{\Gamma}=&\sum_{\left(i,j\right)} \frac{\mathbf{T}_i \cdot \mathbf{T}_j}{2} 
    \gamma_{\text{cusp}} \left(l_{\mu}-2\log{2}+i\pi\right)+\sum_{i} \gamma^i
    \\
    &-\mathbf{T}_Q \cdot \mathbf{T}_{\overline{Q}}\,
    \gamma_{\text{cusp}}{\left(\beta_{Q\overline{Q}}\right)}+2 \gamma^Q +
    (\mathbf{T}_Q+\mathbf{T}_{\overline{Q}}) \cdot \sum_i \mathbf{T}_i\; \gamma_{\text{cusp}} 
    \left(\frac{1}{2}l_{\mu}-\log{2}\right)\\
    =&\sum_{\left(i,j\right)} \frac{\mathbf{T}_i \cdot \mathbf{T}_j}{2} 
    \gamma_{\text{cusp}} \left(l_{\mu}-2\log{2}+i\pi\right)+\sum_{i} \gamma^i\\
    &-\frac{1}{2}\mathbf{T}_\mathcal{Q}^2
    \gamma_{\text{cusp}}{\left(\beta_{Q\overline{Q}}\right)}+2 \gamma^Q+\frac{1}{2}(\mathbf{T}_{Q}^2+\mathbf{T}_{\overline{Q}}^2)\,\gamma_{\text{cusp}}{\left(\beta_{Q\overline{Q}}\right)} -\mathbf{T}_\mathcal{Q}^2 \gamma_{\text{cusp}} 
    \left(\frac{1}{2}l_{\mu}-\log{2}\right)\,,
     \label{eq:anodimmatrix_2}
\end{split}
\end{equation}
where in the last line we introduced the total colour charge $\mathbf{T}_\mathcal{Q} = \mathbf{T}_Q + \mathbf{T}_{\overline{Q}} = -\sum_i \mathbf{T}_i$ of the quarkonium bound state. We have that
\begin{align}
    \begin{split}
    \textbf{T}_{\mathcal{Q}}^2 =& \begin{cases}
    0 & \text{if }\mathcal{Q} \text{ is in colour-singlet state [1]\,,} \\
    C_A & \text{if }\mathcal{Q} \text{ is in colour-octet state [8]\,.}
    \end{cases}
    \label{eq:TQuarkonium}
    \end{split}
\end{align}

At this point we have to address two issues: First, we see that eq.~\eqref{eq:anodimmatrix_2} depends explicitly on the colour charges $\mathbf{T}_{Q}^2$ and $\mathbf{T}_{\overline{Q}}^2$ of the constituent quarks. However, from colour coherence we expect that the structure of the IR divergences is such that the soft gluons do not resolve the short-distance physics, in this case the individual constituent quarks. Instead, the IR structure should only depend on the colour charge $\mathbf{T}_\mathcal{Q}^2$ of the quarkonium bound state. Second, and more strikingly, we work in a NRQCD framework where the heavy quarks are produced at threshold at zero relative velocity, $v=0$. The velocity $v$ is related to the cusp angle $\beta_{Q\overline{Q}}$ through
\beq
\beta_{Q\overline{Q}}=-i\pi+\log{\frac{1+v}{1-v}}\,.
\eeq
Expanding the cusp anomalous dimension around $v=0$, we find
\beq\begin{split}
\gamma_{\text{cusp}}{\left(\beta_{IJ}\right)}
    =&-\frac{i\pi}{2v}\left[\frac{\alpha_s^{\left(n_l\right)}}{\pi} - \left(\frac{\alpha_s^{\left(n_l\right)}}{\pi}\right)^2 C_A{\left(1-\frac{\pi^2}{12}\right)}\right]
    -{2}\widehat{\gamma}^Q+\mathcal{O}{\left(\alpha_s^3,v^1\right)}\,,
\end{split}
\label{eq:gammacuspbetagen}
\end{equation}
where $\widehat{\gamma}^Q = \frac{1}{C_F}\gamma^Q$ is the heavy-quark anomalous dimension with the Casimir scaled out. We see that the cusp anomalous dimension diverges at $v=0$, which corresponds to the Coulomb singularity of the amplitude.  In our NRQCD framework, we put $v=0$ at the integrand level, and the Coulomb divergence manifests itself as poles in the dimensional regulator, captured by the factor $\mathbf{Z}_{\text{Coul.}}$ in eq.~\eqref{eq:defFinRem}. The formula for the soft anomalous dimension in eq.~\eqref{eq:anodimmatrix}, however, is valid in a framework where $v$ is not put to zero from the start. We should therefore start from a variant of eq.~\eqref{eq:anodimmatrix} with $\gamma_{\text{cusp}}{\left(\beta_{IJ}\right)}$ replaced by a constant $\gamma_{\text{cusp}}^{\textrm{thres}}$. In the following we argue what the correct form of the soft anomalous dimension in our framework is, and we determine $\gamma_{\text{cusp}}^{\textrm{thres}}$ through two loops.

To understand this point, let us first discuss the process  
$\gamma \gamma \leftrightarrow {^1S_0^{[1]}}$. Given that the external particles
are either photons or massive quarks, there cannot be any collinear divergences in this channel. In addition, since the heavy quarks are in a colour-singlet state, colour-coherence implies that there are no IR divergences. We must then have that $\mathbf{\Gamma}^{{^1S_0^{[1]}}}_{\gamma \gamma}=0$. 
From \cref{eq:anodimmatrix} with $\gamma_{\text{cusp}}{\left(\beta_{IJ}\right)} \to \gamma_{\text{cusp}}^{\textrm{thres}}$, however, we find that
\begin{equation}
   0 \stackrel{!}{=} \mathbf{\Gamma}^{{^1S_0^{[1]}}}_{\gamma \gamma}=
    C_F \gamma_{\text{cusp}}^{\textrm{thres}}+2C_F\widehat\gamma^Q + \mathcal{O}(\alpha_s^3)\,,
\end{equation}
where we used $\sum_{\left(I, J\right)} \mathbf{T}_I \cdot \mathbf{T}_J=-2 \mathbf{T}_{Q}^2$ with $ \mathbf{T}_{Q}^2=\mathbf{T}_{\overline{Q}}^2=C_F$, and so we find
\begin{equation}
\begin{split}
        \gamma_{\text{cusp}}^{\textrm{thres}}=&-{2}\widehat \gamma^Q + \mathcal{O}{\left(\alpha_s^3\right)}\,.
\end{split}        
\label{eq:gammacuspbetathreshold}
\end{equation}
We note that \cref{eq:gammacuspbetathreshold} is equivalent to
 \cref{eq:gammacuspbetagen}, when simply removing the poles in $v$,
as is customary within the NRQCD framework \cite{Bodwin:1994jh}. Inserting this relation into eq.~\eqref{eq:anodimmatrix_2}, we find
\begin{equation}
\begin{split}
    \mathbf{\Gamma}=&\sum_{\left(i,j\right)} \frac{\mathbf{T}_i \cdot \mathbf{T}_j}{2} 
    \gamma_{\text{cusp}} \left(l_{\mu}-2\log{2}+i\pi\right)+\sum_{i} \gamma^i+\mathbf{T}_\mathcal{Q}^2
    \widehat\gamma^Q-\mathbf{T}_\mathcal{Q}^2 \gamma_{\text{cusp}} 
    \left(\frac{1}{2}l_{\mu}-\log{2}\right)\,.
     \label{eq:anodimmatrix_3}
\end{split}
\end{equation}
Equation~\eqref{eq:anodimmatrix_3} is the soft anomalous dimension matrix that describes the IR singularities of  quarkonium production and decay at two loops. Note in particular that eq.~\eqref{eq:anodimmatrix_3} only depends on the colour charges of the massless external legs and the quarkonium bound state, as expected from colour coherence.

\subsection{Coulomb singularity} \label{sec:coulomb}

We now turn to the second type of singularities that remain in the renormalised amplitudes, namely the Coulomb singularities that are related to the bound state of the heavy quarks and are governed by $\mathbf{Z}_{\text{Coul.}}$. We work in the \msbar-scheme and we define:
\begin{align}
    \mathbf{Z}_{\text{Coul.}}=& 1+\frac{1}{4\epsilon}\left(\frac{\alpha_s^{\left(n_l\right)}}{\pi}\right)^2 \gamma^{p, c}_{\text{Coulomb}} +\mathcal{O}{\left(\alpha_s^3\right)}\,,
\end{align}
where $\gamma^{p, c}_{\text{Coulomb}}$ is the anomalous dimension for the Coulomb singularity. Note that a priori the value of the anomalous dimension may depend on the channel $p$ and the colour $c$. The Coulomb singularity for the colour-singlet form factor $\gamma \gamma \leftrightarrow {^1S_0^{[1]}}$ 
was already known~\cite{Czarnecki:2001zc,Kniehl:2006qw,Feng:2015uha}, 
\begin{equation}\label{eq:gamma_Coul._aa}
\gamma_{\text{Coulomb}}^{\gamma \gamma, {^1S_0^{[1]}}}=-\pi^2\left(C_F^2+\frac{1}{2}C_F C_A\right)\,.
\end{equation}
The other anomalous dimensions have not been considered before and will 
be presented for the first time in this section. We note nevertheless that the authors
of ref.~\cite{Feng:2017hlu} give compelling evidence for the fact that the Coulomb singularity
in $g g \leftrightarrow {^1S_0^{[1]}}$ is the same as in $\gamma \gamma \leftrightarrow {^1S_0^{[1]}}$.

For all form factors we have considered, we find that
\begin{equation}
    \mathbf{Z}_{\text{IR}}^{-1}\, \mathcal{\overline{F}}_{p,c}
    =\left(\frac{\alpha_s^{\left(n_l\right)}}{\pi}\right)^{2}\frac{1}{4\epsilon}
    \gamma^{p, c}_{\text{Coulomb}}+\mathcal{O}{\left(\epsilon^0, \alpha_s^3\right)}\,.
    \label{eq:coulombZ2}
\end{equation}
For the case $(p,c) = (\gamma \gamma, {^1S_0^{[1]}})$ we reproduce eq.~\eqref{eq:gamma_Coul._aa}. For $(p,c) = (gg, {^1S_0^{[1]}})$, we find the same result as in eq.~\eqref{eq:gamma_Coul._aa}, i.e., we find that the anomalous dimension only depends on the colour, but not on the channel:
\begin{equation}
    \gamma_{\text{Coulomb}}^{{^1S_0^{[1]}}}\equiv\gamma_{\text{Coulomb}}^{\gamma \gamma, {^1S_0^{[1]}}}=
    \gamma_{\text{Coulomb}}^{gg, {^1S_0^{[1]}}}\,.
\end{equation}
This is in agreement with refs.~\cite{Czarnecki:2001zc,Kniehl:2006qw,Feng:2015uha} and confirms
the observation made in ref.~\cite{Feng:2017hlu}.
For the colour-octet form factors we find
\begin{equation}
    \gamma_{\text{Coulomb}}^{{^1S_0^{[8]}}}\equiv\gamma_{\text{Coulomb}}^{\gamma g, {^1S_0^{[8]}}}=
    \gamma_{\text{Coulomb}}^{gg, {^1S_0^{[8]}}}=-\pi^2\left(C_F^2-\frac{1}{2}C_F C_A\right)\,.
\end{equation}
We observe that the Coulomb singularity for the colour-octet states differs 
only in the sign of the non-abelian coefficient $C_F C_A$ from the colour-singlet case. We remark that in the QED limit, $C_F\rightarrow 1$, $C_A\rightarrow 0$, $T_F\rightarrow 1$, we reproduce the Coulomb singularity encountered in the para-positronium decay to two photons \cite{Abreu:2022vei,Czarnecki:1999ci}.

To conclude, we find that all our results have the expected IR and Coulomb singularity structure. This is a very strong check of the correctness of our results.


\section{Form factors} \label{sec:formfac}

In this section we present our results for the finite remainder of the form factors for the processes $\gamma \gamma \leftrightarrow {^1S_0^{[1]}}$, $gg \leftrightarrow {^1S_0^{[1]}}$, $\gamma g \leftrightarrow {^1S_0^{[8]}}$ and $gg \leftrightarrow {^1S_0^{[8]}}$. We write the one-loop and two-loop corrections in the following form:
\begin{equation}
    \mathcal{F}_{p,c}^{\text{fin}}=\left(\frac{\alpha_s^{\left(n_l\right)}}{\pi}\right)^q \left[1+\left(\frac{\alpha_s^{\left(n_l\right)}}{\pi}\right) \mathcal{F}_{p,c}^{\text{fin}, (1)}+\left(\frac{\alpha_s^{\left(n_l\right)}}{\pi}\right)^2 \mathcal{F}_{p,c}^{\text{fin}, (2)}\right]+\mathcal{O}{\left(\alpha_s^3\right)},
    \label{eq:finiteamplitude}
\end{equation}
where $\mathcal{F}_{p,c}^{\text{fin}, (1)}$ is the one-loop correction and $\mathcal{F}_{p,c}^{\text{fin}, (2)}$ is the two-loop correction. In the following we will separate the terms that depend on the renormalisation scale $\mu_R$, the factorisation scale $\mu_F$ and the NRQCD scale $\mu_{\Lambda}$ from the scale-independent terms.
Our results can be found in a set of ancillary files which can be obtained from
ref.~\cite{formFactorsGit}.

At one-loop level, we can express the finite remainder as
\begin{equation}
    \mathcal{F}_{p,c}^{\text{fin}, (1)}=\mathcal{F}_{p,c;\text{reg}}^{\text{fin}, (1)}+\mathcal{C}^{(1)}_{\mu_R}+\mathcal{C}^{(1)}_{\mu_F}\,,
\end{equation}
where $\mathcal{C}^{(1)}_{\mu}$ encapsulates the $\mu$-scale dependence at one-loop and is given by
\begin{align}
	\mathcal{C}^{(1)}_{\mu_R}=&\frac{q}{4}\beta_0\, l_{\mu_R}\,,
	\\
    \mathcal{C}^{(1)}_{\mu_F}=&-\frac{q\, C_A}{4} l_{\mu_F}^2 - \frac{1}{4}\left(\tilde{\mathcal{B}}+q\beta_0\right) l_{\mu_F}\,,
\end{align}
where, for convenience, we have defined the quantity
\begin{equation}
	\tilde{\mathcal{B}}=- 4\, q\,C_A \log{2} + \mathbf{T}_{\mathcal{Q}}^2 - i\pi \left(\mathbf{T}_{\mathcal{Q}}^2-2\,q\,C_A\right),
\end{equation}
and $q$ is defined below \cref{eq:atildepc}.

The situation is more involved at two loops. We can express $\mathcal{F}_{p,c}^{\text{fin}, (2)}$ in terms of their different contributions as
\begin{equation}
\begin{split}
\mathcal{F}_{p,c}^{\text{fin}, (2)}=&\mathcal{F}_{p,c;\text{reg}}^{\text{fin}, (2)}+\mathcal{F}^{\text{fin},(2)}_{p,c;\text{lbl}}+\mathcal{F}^{\text{fin},(2)}_{p,c;\text{vac}}+ \mathcal{F}^{\text{fin},(1)}_{p,c;\text{reg}} \,\left(\mathcal{D}^{(1)}_{\mu_R}+\mathcal{C}^{(1)}_{\mu_F}\right)
\\
& + \mathcal{C}^{(2)}_{\mu_R}+\mathcal{C}^{(2)}_{\mu_F}+\mathcal{C}^{(2)}_{\mu_{\Lambda}}+\mathcal{D}^{(1)}_{\mu_R}\, \mathcal{C}^{(1)}_{\mu_F},
\end{split}
\label{eq:twoloopfiniteremainderstru}
\end{equation}
where the scale-dependent terms are given by
\begin{align}
	\mathcal{D}^{(1)}_{\mu_R}=&\frac{1}{4}\left(1+q\right)\beta_0\, l_{\mu_R},
	\\
	\mathcal{C}^{(2)}_{\mu_R}=&\frac{1}{32}q\left(1+q\right)\beta_0^2\,l_{\mu_R}^2+\frac{1}{16}q\,\beta_1\, l_{\mu_R},
	\\
	\begin{split}
	\mathcal{C}^{(2)}_{\mu_F}=&\frac{1}{32}q^2 C_A^2\, l_{\mu_F}^4 + \frac{1}{16}q\, C_A \left(\tilde{\mathcal{B}}+\frac{1}{3}\left(2+3q\right)\beta_0\right) l_{\mu_F}^3
	\\
	&\hspace{-1cm} + \frac{1}{32}\left(\tilde{\mathcal{B}}^2+q\left(1+q\right)\beta_0^2+\left(1+2q\right)\beta_0\, \tilde{\mathcal{B}}-8q\,C_A\, \gamma_{\text{cusp}}^{(1)}\right) l_{\mu_F}^2
	\\
	&\hspace{-1cm} + \left(\gamma_{\text{cusp}}^{(1)} \left(q\,C_A \log{2}+\frac{1}{4}i\pi \left(\mathbf{T}_{\mathcal{Q}}^2-2\,q\,C_A\right)\right)+\frac{1}{2}\mathbf{T}_{\mathcal{Q}}^2\, \widehat{\gamma}^{Q, (1)}+q\, \gamma^{g,(1)}\right) l_{\mu_F}, 
	\end{split}
	\\
	\mathcal{C}^{(2)}_{\mu_{\Lambda}}=&\, \frac{1}{2}\,\gamma_{\text{Coulomb}}\, l_{\mu_{\Lambda}}.
\end{align}
The expressions for $\gamma_{\text{cusp}}^{(1)}$, $\gamma^{g,(1)}$ and $\widehat{\gamma}^{Q,{(1)}}$ refer to the coefficients of the $\mathcal{O}{\left(\alpha_s^2\right)}$ terms given in appendix~\ref{sec:IRRenCoef}.

In eq.~\eqref{eq:twoloopfiniteremainderstru}, $\mathcal{F}_{p,c;\text{reg}}^{\text{fin}, (2)}$ is the regular contribution, while $\mathcal{F}^{\text{fin},(2)}_{p,c;\text{lbl}}$ represents the light-by-light contributions and $\mathcal{F}^{\text{fin},(2)}_{p,c;\text{vac}}$ is the contribution due to vacuum polarisation diagrams. These contributions can be further decomposed as
\begin{align}
    \mathcal{F}_{p,c;\text{reg}}^{\text{fin}, (2)}=&C_F^2\, a^{(2)}_{p,c;FF} + C_F C_A\, a^{(2)}_{p,c;FA} + C_A^2\, a^{(2)}_{p,c;AA}\,,
    \label{eq:F2regcontrgen}
    \\
    \mathcal{F}^{\text{fin},(2)}_{p,c;\text{lbl}}=&C_F T_F n_h\, b^{(2)}_{p,c;Fh} + C_F T_F \tilde{n}_l\, b^{(2)}_{p,c;Fl} + C_A T_F n_h\, b^{(2)}_{p,c;Ah} + C_A T_F \tilde{n}_l\, b^{(2)}_{p,c;Al}\,,
    \\
    \mathcal{F}^{\text{fin},(2)}_{p,c;\text{vac}}=&C_F T_F n_h\, c^{(2)}_{p,c;Fh} + C_F T_F n_l\, c^{(2)}_{p,c;Fl} + C_A T_F n_h\, c^{(2)}_{p,c;Ah} + C_A T_F n_l\, c^{(2)}_{p,c;Al}\,.
    \label{eq:F2vaccontrgen}
\end{align}
The definition to the quantity $\tilde{n}_l$ can be found in eq.~\eqref{eq:definitionnltilde}.
The purely abelian contributions at one-loop and two-loop level for the regular and vacuum polarisation corrections are identical for all form factors considered, which is a strong check of the calculation. As mentioned earlier, the abelian contribution for the light-by-light contributions depends on the colour state of the bound state, but is independent of the initial-state partons. Its value differs only by a factor of two between colour-singlet and colour-octet states and comes from the different colour algebra. Similarly, for the non-abelian light-by-light contributions, we can reconstruct the contributions to the colour-octet form factors using the results obtained in the colour-singlet case by simple colour algebra and we find again full agreement, which is another strong check of the calculation.

For phenomenological applications we are interested in the hard functions
obtained by squaring the form factors we obtain in this paper. More explicitly,
the hard function $\mathcal{H}$ is defined as
\begin{equation}
\begin{split}
    \mathcal{H}_{p,c}=&\left\vert \sum_{i=0} \left(\frac{\alpha_s^{\left(n_l\right)}}{\pi}\right)^{i} \mathcal{F}_{p,c}^{\text{fin}, (i)} \right\vert^2,
    \\
    =& \mathcal{H}_{p,c}^{(0)} + \left(\frac{\alpha_s^{\left(n_l\right)}}{\pi}\right) \mathcal{H}_{p,c}^{(1)} + \left(\frac{\alpha_s^{\left(n_l\right)}}{\pi}\right)^2 \mathcal{H}_{p,c}^{(2)}+\mathcal{O}{\left(\alpha_s^3\right)},
\end{split}    
\end{equation}
where we have that $\mathcal{H}_{p,c}^{(0)}=1$ due to the normalisation 
of $\mathcal{F}_{p,c}^{\text{fin},(0)}=1$. In the following subsections, we
present our results for $\mathcal{H}$ for the different channels we consider
in this paper.

Before doing so, however, we close this introductory discussion by noting that by
taking the QED limit, $C_F\rightarrow 1$, $C_A\rightarrow 0$, $T_F\rightarrow 1$, of a colour-singlet form factor, and setting $n_l=0$ we reproduce the form factor of the para-positronium decay into two photons~\cite{Czarnecki:1999ci} which we have already presented in our companion paper \cite{Abreu:2022vei}. The form factors can also be used to describe the para-muonium and para-tauonium decay into two photons up to NNLO accuracy. In this case, one again takes the QED limit but now retains the $n_l$ term. Indeed, in addition to the light fermions ($e$ for muonium and $e, \mu$ for tauonium), one now also has to take into account light quarks and their relative charge. We will denote generically the relative charge of leptons and quarks that form the bound state as $e_f$. If we wish to compute the NNLO QED corrections to quarkonium or leptonium decay into two photons, we would need to make the following replacements $\alpha_s \rightarrow e_f^2 \alpha_{em}$, $\tilde{n}_l\rightarrow n_l^{\text{lbl,QED}}$, $n_l\rightarrow n_l^{\text{vac,QED}}$, with
\begin{equation}
    n_l^{\text{lbl,QED}}=\sum_i^{n_l} e_i^4/e_f^4\,,\quad\quad\quad\quad n_l^{\text{vac,QED}}=\sum_i^{n_l} e_i^2/e_f^2\,.
\end{equation}
For the leptonium bound states, one would consider also the non-perturbative effects from the bound state, including the removal of the Coulomb singularity, similarly to what was done for the para-positronium case in refs.~\cite{Abreu:2022vei,Czarnecki:1999ci}.

\subsection{Form factor coefficients}

In this subsection we present the set of independent coefficients that appear in the different form factors. Since the analytical expressions are rather lengthy, we have collected the complete analytical expressions for all coefficients expressible in terms of multiple polylogarithms, elliptic multiple polylogarithms and iterated integrals of Eisenstein series in appendix~\ref{sec:formfactoranalyticexpression} and in a set
of ancillary files that can be obtained from ref.~\cite{formFactorsGit}.
We also have available  high-precision numerical evaluations up to more than 1000 digits. In the following, we will show only the first 20 digits of the numerical evaluations.

For the one-loop coefficients, we define the following coefficients,
\begin{align}
	a_{1}^{(1)}=&\frac{\pi^2}{8}-\frac{5}{2}=-1.2662994498638301726,
	\\
	a_{2}^{(1)}=&\frac{\pi^2}{6}+\frac{1}{2}-\log^2{2}+i\pi \log{2}=1.66448105293002501181 + i\, 2.17758609030360213050,
	\\
	a_{3}^{(1)}=&\frac{\pi^2}{48}+\frac{3}{4}+\frac{1}{2}\log{2}-\frac{1}{2}\log^2{2}=1.0619638416769002469.
\end{align}
At two-loop level, we define
\begin{align}
	a_{1}^{(2)}=&a^{(2)}_{\gamma \gamma,[1];FF}=-21.10789796731067145661,
	\\
	a_{2}^{(2)}=&a^{(2)}_{\gamma \gamma,[1];FA}=-4.79298000108431445013,
	\\
	a_{3}^{(2)}=&a^{(2)}_{gg,[1];FA}=-1.63396444740133643183 - i\, 2.75747606818258018891,
	\\
	a_{4}^{(2)}=&a^{(2)}_{\gamma g,[8];FA}=11.4964197929416576889,
	\\
	a_{5}^{(2)}=&a^{(2)}_{gg,[1];AA}=-4.16141057462231200330 + i\, 12.74963942099565970837,
	\\
	a_{6}^{(2)}=&a^{(2)}_{\gamma g,[8];AA}=1.1674088877410300338,
	\\
	a_{7}^{(2)}=&a^{(2)}_{g g,[8];AA}= -0.47052470943276749673 + i\, 3.76949207800965060010,
	\\
	b_{1}^{(2)}=&b^{(2)}_{\gamma \gamma,[1];Fh}=0.64696557211233073992 + i\, 2.07357555846158085167,
	\\
	b_{2}^{(2)}=&b^{(2)}_{\gamma \gamma,[1];Fl}=0.73128459201956765416 - i\, 1.79084590261634204461,
	\\
	b_{3}^{(2)}=&b^{(2)}_{gg,[1];Ah}= 0.17355457403625922073 + i\, 0.27096443988081661998,
	\\
	b_{4}^{(2)}=&b^{(2)}_{gg,[1];Al}= -0.27562418279938901511 + i\, 0.56534858757600288424,
	\\
	c_{1}^{(2)}=&c^{(2)}_{\gamma \gamma,[1];Fh}=0.22367201327357266787,
	\\
	c_{2}^{(2)}=&c^{(2)}_{\gamma \gamma,[1];Fl}=-0.56481511444874563705,
	\\
	c_{3}^{(2)}=&c^{(2)}_{gg,[1];Ah}= -0.19435982593932209621 - i\, 0.26424642484869050250,
	\\
	c_{4}^{(2)}=&c^{(2)}_{\gamma g,[8];Ah}= -0.1068025969267476488,
	\\
	c_{5}^{(2)}=&c^{(2)}_{gg,[1];Al}= 0.20360900095614056680 - i\, 2.96547152392125649208,
	\\
	c_{6}^{(2)}=&c^{(2)}_{\gamma g,[8];Al}= -0.5846981879646550889.
\end{align}

\subsection{$\gamma \gamma \leftrightarrow {^1S_0^{[1]}}$}

For the form factor $\mathcal{F}_{\gamma \gamma, [1]}^{\text{fin}}$, the correction at one-loop accuracy reads
\begin{equation}
\begin{split}
    \mathcal{F}_{\gamma \gamma, [1]; 
    \text{reg}}^{\text{fin},(1)}=&C_F\, a_{1}^{(1)}.
\end{split}
\end{equation}
At two-loop level, the individual coefficients in eqs.~\eqref{eq:F2regcontrgen}-\eqref{eq:F2vaccontrgen} read
\begin{gather}
   \hspace{-0.1cm} a^{(2)}_{\gamma \gamma,[1];FF}=a_{1}^{(2)}, \qquad\qquad\quad\;\;    a^{(2)}_{\gamma \gamma,[1];FA}=a_{2}^{(2)}, \qquad\qquad\quad\;\;   a^{(2)}_{\gamma \gamma,[1];AA}=0, \nonumber
    \\[5pt]
    b^{(2)}_{\gamma \gamma,[1];Fh}=b_1^{(2)}, \qquad    b^{(2)}_{\gamma \gamma,[1];Fl}=b_2^{(2)}, \qquad   b^{(2)}_{\gamma \gamma,[1];Ah}=0, \qquad    b^{(2)}_{\gamma \gamma,[1];Al}=0,
    \\[5pt]
    c^{(2)}_{\gamma \gamma,[1];Fh}=c_1^{(2)}, \qquad    c^{(2)}_{\gamma \gamma,[1];Fl}=c_2^{(2)}, \qquad    c^{(2)}_{\gamma \gamma,[1];Ah}=0, \qquad    c^{(2)}_{\gamma \gamma,[1];Al}=0. \nonumber
\end{gather}
As mentioned before, the coefficients for the regular and vacuum insertion contributions have been computed in numerical form for the first time in ref.~\cite{Czarnecki:2001zc}. The light-by-light contribution has been considered in ref.~\cite{Feng:2015uha} where all coefficients have been evaluated at an improved numerical precision of 10 digits. We find full agreement for all the coefficients presented in both references.

Using the results for this form factor, we can compute the hard function that can be used for the decay width into two photons. We obtain
\begin{align}
    \mathcal{H}_{\gamma \gamma,[1]}^{(1)}=&-3.37679853297021379372,
    \\
    \begin{split}
    \mathcal{H}_{\gamma \gamma,[1]}^{(2)}=&-109.3826016955304736674 - 9.2861959656680879327\, l_{\mu_R}
    \\
    &- 37.2851721818931325600\, l_{\mu_{\Lambda}} - 0.7530868192649941827\, n_l
    \\
    & + 0.5627997554950356323\, n_l \, l_{\mu_R} + 0.97504612269275687222\, \tilde{n}_l.
    \end{split}
\end{align}
For charmonium decay, we set $n_l=3$ and in the bottomonium case $n_l=4$. The light-by-light contributions which have been omitted in ref.~\cite{Czarnecki:2001zc} contain the term $\tilde{n}_l$ which turns out to be quite large for the bottomonium state
\begin{equation}
    \tilde{n}_l=\sum_i^{n_l} \frac{e_i^2}{e_Q^2}= \begin{cases} 3/2 & \text{for }c\overline{c}, \\ 10 & \text{for }b\overline{b}. \end{cases}
\end{equation}

\subsection{$gg\leftrightarrow {^1S_0^{[1]}}$}

In this subsection, we present for the first time the form factor $\mathcal{F}_{g g, [1]}^{\text{fin}}$. The correction at one-loop accuracy reads
\begin{equation}
\begin{split}
    \mathcal{F}_{gg, [1];\text{reg}}^{\text{fin},(1)}=&C_F \, a_{1}^{(1)}+C_A \, a_{2}^{(1)}.
\end{split}
\end{equation}
The two-loop coefficients read
\begin{gather}
   \hspace{-1cm} a^{(2)}_{gg,[1];FF}=a_{1}^{(2)}, \qquad\qquad\quad\;\;\;\,    a^{(2)}_{gg,[1];FA}=a_{3}^{(2)}, \qquad\qquad\quad\;\;\;\,    a^{(2)}_{gg,[1];AA}=a_{5}^{(2)}, \nonumber
    \\[5pt]
    b^{(2)}_{gg,[1];Fh}= b_{1}^{(2)}, \qquad    b^{(2)}_{gg,[1];Fl}= b_{2}^{(2)}, \qquad    b^{(2)}_{gg,[1];Ah}= b_{3}^{(2)}, \qquad    b^{(2)}_{gg,[1];Al}= b_{4}^{(2)},
    \\[5pt]
   \hspace{-0.8cm} c^{(2)}_{gg,[1];Fh}= c_{1}^{(2)}, \qquad   c^{(2)}_{gg,[1];Fl}= c_{2}^{(2)}, \qquad   c^{(2)}_{gg,[1];Ah}= c_{3}^{(2)}, \qquad    c^{(2)}_{gg,[1];Al}= c_{5}^{(2)}. \nonumber
\end{gather}

The hard function, which can be used in collinear or Transverse-Momentum-Dependent (TMD) factorisation, exhibits the following structure:
\begin{align}
    \begin{split}
    \mathcal{H}_{gg,[1]}^{(1)}=& 6.6100877846099362771 + 5.5000000000000000000\, l_{\mu_R}
    \\
    &-1.3411169166403281435\, l_{\mu_F} - 1.50000000000000000000\, l_{\mu_F}^2
    \\
    &-0.3333333333333333333\, n_l \left(l_{\mu_R}-l_{\mu_F}\right),
    \end{split}
    \\
    \begin{split}
    \mathcal{H}_{gg,[1]}^{(2)}=& -108.32872969182897851535 + 67.28322422303197428616\, l_{\mu_R}
    \\
    &+22.6875000000000000000\, l_{\mu_R}^2 -15.24945497092079433491\, l_{\mu_F}
    \\
    & -11.8456969740765133031\, l_{\mu_F}^2 +4.76167537496049221524 \, l_{\mu_F}^3
    \\
    & +1.1250000000000000000 \, l_{\mu_F}^4 -11.0642145622827071838 \, l_{\mu_R} \,l_{\mu_F}
    \\
    & -12.3750000000000000000  \, l_{\mu_R} \,l_{\mu_F}^2 - 37.2851721818931325600\, l_{\mu_{\Lambda}}
    \\
    & + 0.0059137578980173446\, n_l -4.88837722563830147189 \, n_l \, l_{\mu_R}
    \\
    & -2.7500000000000000000 \, n_l \, l_{\mu_R}^2 +2.7479948883357910787 \, n_l \, l_{\mu_F}  
	\\
    & -0.6004653819334700598 \, n_l \, l_{\mu_F}^2 -0.66666666666666666667 \, n_l \, l_{\mu_F}^3
    \\
    & + 3.42055845832016407175 \, n_l \, l_{\mu_R} \, l_{\mu_F}  +0.75000000000000000000 \, n_l \, l_{\mu_R} \, l_{\mu_F}^2
    \\
    & + 0.08333333333333333333 \,n_l^2\left(l_{\mu_R}-l_{\mu_F}\right)^2.
    \end{split}
\end{align}
We note that the size of the coefficient multiplying the logarithm of the NRQCD scale is rather large and has an important effect on the numerical value of the hard function.

\subsection{$\gamma g\leftrightarrow {^1S_0^{[8]}}$}

As for the colour-octet states, we first consider the form factor $\mathcal{F}_{\gamma g, [8]}^{\text{fin}}$. At one-loop accuracy, the correction is given by
\begin{equation}
\begin{split}
    \mathcal{F}_{\gamma g, [8]; \text{reg}}^{\text{fin},(1)}=&C_F \, a_{1}^{(1)} + C_A \, a_{3}^{(1)}.
\end{split}
\end{equation}
At two-loop order, the coefficients read,
\begin{gather}
   \hspace{-1.6cm} a^{(2)}_{\gamma g,[8];FF}=a_{1}^{(2)}, \qquad\qquad\qquad\quad\;\,    a^{(2)}_{\gamma g,[8];FA}=a_{4}^{(2)}, \qquad\qquad\qquad\quad\;\,    a^{(2)}_{\gamma g,[8];AA}=a_{6}^{(2)}, \nonumber
    \\[5pt]
    b^{(2)}_{\gamma g,[8];Fh}=2\, b_{1}^{(2)}, \qquad    b^{(2)}_{\gamma g,[8];Fl}=2\, b_{2}^{(2)}, \qquad   b^{(2)}_{\gamma g,[8];Ah}= -\frac{3}{4}\, b_{1}^{(2)}, \qquad   b^{(2)}_{\gamma g,[8];Al}= -\frac{3}{4}\, b_{2}^{(2)},
    \\[5pt]
   \hspace{-1.45cm} c^{(2)}_{\gamma g,[8];Fh}= c_{1}^{(2)}, \qquad\;\;   c^{(2)}_{\gamma g,[8];Fl}= c_{2}^{(2)}, \qquad\;\;\;   c^{(2)}_{\gamma g,[8];Ah}= c_{4}^{(2)}, \qquad\quad\;\;\,   c^{(2)}_{\gamma g,[8];Al}= c_{6}^{(2)}. \nonumber
\end{gather}

The hard function for the colour-octet state in the channel $\gamma g$ takes the form
\begin{align}
    \begin{split}
    \mathcal{H}_{\gamma g,[8]}^{(1)}=& 2.9949845170911876879 + 2.7500000000000000000\, l_{\mu_R}
    \\
    &-2.1705584583201640717\, l_{\mu_F} - 0.75000000000000000000\, l_{\mu_F}^2
    \\
    & - 0.16666666666666666667 n_l \left(l_{\mu_R}-l_{\mu_F}\right),
    \end{split}
    \\
    \begin{split}
    \mathcal{H}_{\gamma g,[8]}^{(2)}=& 40.4242880521358345950 + 22.84741484400153228336\, l_{\mu_R}
    \\
    & + 7.56250000000000000000 \, l_{\mu_R}^2 - 14.8215925690621966374 \, l_{\mu_F}
    \\
    & + 0.75699232806869328971 \, l_{\mu_F}^2 + 3.0029188437401230538 \, l_{\mu_F}^3
    \\
    & + 0.2812500000000000000 \, l_{\mu_F}^4 -11.93807152076090239462 \, l_{\mu_R} \, l_{\mu_F}
    \\
    & -4.1250000000000000000 \, l_{\mu_R} \, l_{\mu_F}^2 + 2.19324542246430191530\, l_{\mu_{\Lambda}}
    \\
    & - 2.5071813831589594496\, n_l -1.78999483903039589596 \, n_l \, l_{\mu_R}
    \\
    & -0.9166666666666666667 \, n_l \, l_{\mu_R}^2 +1.18814689958143744113 \, n_l \, l_{\mu_F}
    \\
    & -0.5634729479133743513 \, n_l \, l_{\mu_F}^2 -0.20833333333333333333 \, n_l \, l_{\mu_F}^3
    \\
    & + 1.64018615277338802392 \, n_l \, l_{\mu_R} \, l_{\mu_F} + 0.2500000000000000000 \, n_l \, l_{\mu_R} \, l_{\mu_F}^2 
    \\
    & + 0.02777777777777777778 \, n_l^2 \left(l_{\mu_R}-l_{\mu_F}\right)^2 + 0.30470191334148652257\, \tilde{n}_l.
    \end{split}
\end{align}
The variable $\tilde{n}_l$ vanishes for charmonium states and takes a negative value for bottomonium states
\begin{equation}
    \tilde{n}_l=\sum_i^{n_l} \frac{e_i}{e_Q}= \begin{cases} 0 & \text{for }c\overline{c}, \\ -2 & \text{for }b\overline{b}. \end{cases}
\end{equation}

\subsection{$gg\leftrightarrow {^1S_0^{[8]}}$}

For the second colour-octet form factor $\mathcal{F}_{g g, [8]}^{\text{fin}}$, the relative correction at one-loop level is
\begin{equation}
\begin{split}
    \mathcal{F}_{g g, [8];\text{reg}}^{\text{fin},(1)}=&C_F \, a_{1}^{(1)} + C_A \left(\frac{1}{2}a_{2}^{(1)}+a_{3}^{(1)}\right).
\end{split}
\end{equation}
At two-loop order the coefficients read,
\begin{gather}
    a^{(2)}_{g g,[8];FF}=a_{1}^{(2)}, \qquad\quad    a^{(2)}_{g g,[8];FA}= -\frac{1}{2}\, a_{2}^{(2)} + \frac{1}{2}\, a_{3}^{(2)} + a_{4}^{(2)}, \qquad\quad    a^{(2)}_{g g,[8];AA}= a_{7}^{(2)}, \nonumber
    \\[5pt]
 \hspace{-1.8cm}   b^{(2)}_{g g,[8];Fh}=2\, b_{1}^{(2)}, \qquad\qquad\qquad\qquad\;\;   b^{(2)}_{g g,[8];Fl}=2\, b_{2}^{(2)} \nonumber,
    \\[5pt]
    b^{(2)}_{g g,[8];Ah}= -\frac{3}{4}\, b_{1}^{(2)} + \frac{1}{2}\, b_{3}^{(2)}, \qquad\qquad   b^{(2)}_{g g,[8];Al}= -\frac{3}{4}\, b_{2}^{(2)} + \frac{1}{2}\, b_{4}^{(2)},
    \\[5pt]
 \hspace{-1.8cm}    c^{(2)}_{g g,[8];Fh}= c_{1}^{(2)}, \qquad\qquad\qquad\qquad\;\;\;\;\;    c^{(2)}_{g g,[8];Fl}= c_{2}^{(2)}, \nonumber
    \\[5pt]
 \hspace{-0.45cm}   c^{(2)}_{g g,[8];Ah}= \frac{1}{2}\, c_{3}^{(2)} + c_{4}^{(2)}, \qquad\qquad\qquad   c^{(2)}_{g g,[8];Al}= \frac{1}{2}\, c_{5}^{(2)} + c_{6}^{(2)}. \nonumber
\end{gather}

The hard function for the second colour-octet state in the $gg$-channel is given by
\begin{align}
    \begin{split}
    \mathcal{H}_{g g,[8]}^{(1)}=& 7.9884276758812627233 + 5.5000000000000000000\, l_{\mu_R}
    \\
    &-2.8411169166403281435 \, l_{\mu_F} - 1.50000000000000000000\, l_{\mu_F}^2
    \\
    & -0.3333333333333333333 \left(l_{\mu_R}-l_{\mu_F}\right),
    \end{split}
    \\
    \begin{split}
    \mathcal{H}_{g g,[8]}^{(2)}=& 47.92683141521562851467 + 78.6545283260204174672 \, l_{\mu_R}
    \\
    & +22.6875000000000000000 \, l_{\mu_R}^2 -34.2091378115074325464 \, l_{\mu_F}
    \\
    & -8.7140314360230107571 \, l_{\mu_F}^2 + 7.01167537496049221524 \, l_{\mu_F}^3
    \\
    & + 1.1250000000000000000 \, l_{\mu_F}^4 -23.4392145622827071838 \, l_{\mu_R} \, l_{\mu_F}
    \\
    & -12.3750000000000000000 \, l_{\mu_R} \, l_{\mu_F}^2 + 2.19324542246430191530 \, l_{\mu_{\Lambda}}
    \\
    & - 2.3105022425823455994 \, n_l -5.5775471712739646950 \, n_l \, l_{\mu_R}
    \\
    & -2.7500000000000000000 \, n_l \, l_{\mu_R}^2 +3.62410818542623322740 \, n_l \, l_{\mu_F}
    \\
    & -1.2254653819334700598 \, n_l \, l_{\mu_F}^2 -0.66666666666666666667 \, n_l \, l_{\mu_F}^3
    \\
    & + 4.1705584583201640717 \, n_l \, l_{\mu_R} \, l_{\mu_F} + 0.75000000000000000000 \, n_l \, l_{\mu_R} \, l_{\mu_F}^2
    \\
    & +0.08333333333333333333 \, n_l^2 \left(l_{\mu_R}-l_{\mu_F}\right)^2.
    \end{split}
\end{align}
Comparing the size of the coefficient of the NRQCD scale dependence of the colour-octet states with the situation in the colour-singlet case, we can conclude that the hard function is not as sensitive to the NRQCD scale as it is in the colour-singlet case.


\section{Conclusions} \label{sec:conclusions}

In this paper we have computed analytically the complete two-loop QCD
corrections to the form factors relevant for $\eta_Q$ production and decay.
In particular, we have considered the processes
$\gamma \gamma \leftrightarrow {^1S_0^{[1]}}$, $gg \leftrightarrow {^1S_0^{[1]}}$, 
$\gamma g \leftrightarrow {^1S_0^{[8]}}$, $gg \leftrightarrow {^1S_0^{[8]}}$.
We have also obtained high-precision numerics up to 1000 digits for
all form factors, which makes our results readily usable for phenomenological studies.
The form factors presented also allow us to consider the
two-loop QED corrections to leptonium bound states.

The form factor $\gamma \gamma \leftrightarrow {^1S_0^{[1]}}$ has been
computed before only in purely numerical form~\cite{Czarnecki:2001zc, Feng:2015uha}. Our result is in agreement with those references, which serves as a cross-check of our calculation. The form factor $gg \leftrightarrow {^1S_0^{[1]}}$ is new and is the last missing ingredient for a full NNLO calculation of $\eta_Q$ hadro-production in either collinear or TMD factorisation. We also computed the form factors to produce a pseudo-scalar state in a colour-octet configuration ${^1S_0^{[8]}}$, which corresponds to higher terms in the $v$-expansion of the LDME. For instance, the pseudo-scalar state ${^1S_0^{[8]}}$ turns out to be one of the leading contributions to the pseudo-vector particle $h_Q$, the other being the state ${^1P_1^{[1]}}$. It also appears in the higher terms in the $v$-expansion for the vector particles $J/\psi$ and $\Upsilon$.

The two-loop bare form factors can be expressed in terms of 76 master
integrals, which we have already discussed in our companion 
paper \cite{Abreu:2022vei}. After UV renormalisation, the renormalised amplitude still contains IR as well as Coulomb singularities.
By imposing that the result has the expected IR pole structure, we were able
to reproduce the Coulomb singularity of the colour-singlet state ${^1S_0^{[1]}}$. 
This serves as a cross-check of our approach. This singularity is
independent of the initial-state particles and depends only on the
bound-state colour configuration. In addition, we obtain for the first time the
Coulomb singularity for the colour-octet state ${^1S_0^{[8]}}$. It differs
only in the non-abelian part from the one in the colour-singlet case.

We have presented the finite remainders for the form factors in 
section~\ref{sec:formfac}. The complete analytical expressions to the coefficients 
can be found in appendix~\ref{sec:formfactoranalyticexpression} 
and in a set of ancillary files \cite{formFactorsGit}. In addition to
this, we have presented the hard function for all processes including the
dependence on the renormalisation scale $\mu_R$, the factorisation scale
$\mu_F$ and the NRQCD scale $\mu_{\Lambda}$. These hard functions can now be
directly used for phenomenology, e.g., when computing the NNLO
corrections to $\eta_Q$ hadro-production. We leave this for future work.

\acknowledgments{We acknowledge useful discussions with Jean-Philippe Lansberg, Kirill Melnikov,  Hua-Sheng Shao and Robert Szafron.
M.B.~acknowledges the financial support from the European Union Horizon 
2020 research and innovation programme: High precision multi-jet dynamics 
at the LHC (grant agreement no. 772009). 
The research of C.D.~and M.A.O.~was supported by the ERC Starting Grant 637019 ``MathAm''. 
M.A.O.~thanks the TH Department at CERN for hospitality while part of this work was carried out.
M.A.O.~also acknowledges the financial support from the following sources: the European Union's Horizon 2020 research and innovation programme under grant agreement STRONG'2020 No 824093 in order to contribute to the EU Virtual Access NLOAccess (VA1-WG10), the funding from the Agence Nationale de la Recherche (ANR) via the grant ANR-20-CE31-0015 (``PrecisOnium'') and via the IDEX Paris-Saclay ``Investissements d'Avenir'' (ANR-11-IDEX-0003-01) through the GLUODYNAMICS project funded by the ``P2IO LabEx (ANR-10-LABX-0038)'', and partially via the IN2P3 project GLUE@NLO funded by the French CNRS.
}

\appendix

\newpage

\section{Bare amplitude structure}
\label{sec:barepole}

In this appendix, we give the structure of the bare amplitude for each form factor at one-loop level up to $\mathcal{O}{\left(\epsilon^2\right)}$ and at two-loop level up to the finite piece $\mathcal{O}{\left(\epsilon^0\right)}$. The decomposition of the bare amplitude follows the notation in section~\ref{sec:bareamp}.

For the one-loop amplitude, we furnish the analytic expressions up to $\mathcal{O}{\left(\epsilon^1\right)}$, while for the highest order term we give its numerical value.  In the case of the two-loop amplitude, we provide the analytic expressions for the pole structure, while for the finite piece we furnish the numerical value. For convenience, we display only the first $5$ digits after the decimal.

At one-loop order we have the following results for the different form factors:

\vspace{0.3cm}
$\gamma \gamma \leftrightarrow {^1S_0^{[1]}}$:
\begin{center}
\renewcommand{\arraystretch}{1.5}
\begin{tabular}{| m{1cm} || m{1cm} | m{2.0cm} | m{2.5cm} | m{4.0cm} | m{1.5cm} |} 
 \hline
 $\gamma \gamma, [1]$ & $\epsilon^{-2}$ & $ \epsilon^{-1}$ & $\epsilon^{0}$ & $\epsilon^{1}$ & $\epsilon^{2}$ \\[5pt]
 \hline\hline
 $\mathcal{F}^{(1)}_F$ & 0 & 0 & $\frac{\pi^2}{8}-\frac{5}{2}$ & $-1+\frac{1}{4}\pi^2+4 \log{2}+\frac{7}{4} \zeta_3$ & $5.52395$ \\[5pt]
 \hline
 $\mathcal{F}^{(1)}_A$ & 0 & 0 & 0 & 0 & 0 \\[5pt]
 \hline
\end{tabular}
\end{center}

\vspace{0.1cm}
$gg \leftrightarrow {^1S_0^{[1]}}$:
\begin{center}
\renewcommand{\arraystretch}{1.5}
\begin{tabular}{| m{1cm} || m{1cm} | m{2.0cm} | m{2.5cm} | m{4.0cm} | m{1.5cm} |} 
 \hline
 $gg, [1]$ & $\epsilon^{-2}$ & $ \epsilon^{-1}$ & $\epsilon^{0}$ & $\epsilon^{1}$ & $\epsilon^{2}$ \\[5pt]
 \hline\hline
 $\mathcal{F}^{(1)}_F$ & 0 & 0 & $\frac{\pi^2}{8}-\frac{5}{2}$ & $-1+\frac{1}{4}\pi^2+4 \log{2}+\frac{7}{4} \zeta_3$ & $5.52395$ \\[5pt]
 \hline
 $\mathcal{F}^{(1)}_A$ & $-\frac{1}{2}$ & $-\frac{i\pi}{2}+\log{2}$ & $\frac{1}{2}+\frac{1}{6}\pi^2+i \pi \log{2}-\log^2{2}$ & $1+i \pi+\frac{1}{8}\pi^2+\frac{i}{8} \pi^3-2 \log{2}-\frac{7}{12} \pi^2 \log{2}-i \pi \log^2{2}+\frac{2}{3} \log^3{2}-\frac{7}{12} \zeta_3$ & $-9.59291 + i\, 4.79988$ \\[5pt]
 \hline
\end{tabular}
\end{center}

\vspace{0.1cm}
$\gamma g \leftrightarrow {^1S_0^{[8]}}$:
\begin{center}
\renewcommand{\arraystretch}{1.5}
\begin{tabular}{| m{1cm} || m{1cm} | m{2.0cm} | m{2.5cm} | m{4.0cm} | m{1.5cm} |} 
 \hline
 $\gamma g, [8]$  & $\epsilon^{-2}$ & $ \epsilon^{-1}$ & $\epsilon^{0}$ & $\epsilon^{1}$ & $\epsilon^{2}$ \\[5pt]
 \hline\hline
 $\mathcal{F}^{(1)}_F$ & 0 & 0 & $\frac{\pi^2}{8}-\frac{5}{2}$ & $-1+\frac{1}{4}\pi^2+4 \log{2}+\frac{7}{4} \zeta_3$ & $5.52395$ \\[5pt]
 \hline
 $\mathcal{F}^{(1)}_A$ & $-\frac{1}{4}$ & $-\frac{1}{4}+\frac{1}{2}\log{2}$ & $\frac{3}{4}+ \frac{1}{48}\pi^2+\frac{1}{2}\log{2}-\frac{1}{2}\log^2{2}$ & $-\frac{1}{2}-\frac{1}{6}\pi^2-\frac{1}{2}\log{2}-\frac{1}{24} \pi^2 \log{2}-\frac{1}{2}\log^2{2}+\frac{1}{3}\log^3{2}+\frac{7}{12} \zeta_3$ & $-1.12737$ \\[5pt]
 \hline
\end{tabular}
\end{center}

\vspace{0.1cm}
$gg \leftrightarrow {^1S_0^{[8]}}$:
\begin{center}
\renewcommand{\arraystretch}{1.5}
\begin{tabular}{| m{1cm} || m{1cm} | m{2.0cm} | m{2.5cm} | m{4.0cm} | m{1.5cm} |} 
 \hline
 $gg, [8]$ & $\epsilon^{-2}$ & $ \epsilon^{-1}$ & $\epsilon^{0}$ & $\epsilon^{1}$ & $\epsilon^{2}$ \\[5pt]
 \hline\hline
 $\mathcal{F}^{(1)}_F$ & 0 & 0 & $\frac{\pi^2}{8}-\frac{5}{2}$ & $-1+\frac{1}{4}\pi^2+4 \log{2}+\frac{7}{4} \zeta_3$ & $5.52395$ \\[5pt]
 \hline
 $\mathcal{F}^{(1)}_A$ & $-\frac{1}{2}$ & $-\frac{i\pi}{4}-\frac{1}{4}+\log{2}$ & $1+\frac{5}{48} \pi^2+\frac{1}{2}\log{2}+\frac{1}{2} i \pi \log{2}-\log^2{2}$ & $\frac{i}{2} \pi-\frac{5}{48} \pi^2+\frac{i}{16} \pi^3-\frac{3}{2} \log{2}-\frac{1}{3} \pi^2 \log{2}-\frac{1}{2}\log^2{2}-\frac{1}{2} i \pi \log^2{2}+\frac{2}{3} \log^3{2}+\frac{7}{24} \zeta_3$ & $-5.92383+i\, 2.39994 $ \\[5pt]
 \hline
\end{tabular}
\end{center}

At two-loop level we find the following structures for the bare amplitudes:

\vspace{0.3cm}
$\gamma \gamma \leftrightarrow {^1S_0^{[1]}}$:
\begin{center}
\renewcommand{\arraystretch}{1.5}
\begin{tabular}{| m{1.1cm} || m{0.7cm} | m{2cm} | m{2.5cm} | m{4.5cm} | m{1.5cm} |}  
 \hline
  $\gamma \gamma, [1]$ & $\epsilon^{-4}$ & $\epsilon^{-3}$ & $\epsilon^{-2}$ & $\epsilon^{-1}$ & $\epsilon^{0}$ \\[5pt]
 \hline\hline
 $\mathcal{F}^{(2)}_{FF}$ & 0 & 0 & $\frac{3}{32}$ & $-\frac{39}{32}-\frac{\pi^2}{16}+\frac{3}{4}\log{2}$ & $-9.58245$ \\[5pt]
 \hline
 $\mathcal{F}^{(2)}_{FA}$ & 0 & 0 & 0 & $-\frac{205}{96}-\frac{\pi^2}{96}$ & 1.56657 \\[5pt]
 \hline
 $\mathcal{F}^{(2)}_{AA}$ & 0 & 0 & 0 & 0 & 0 \\[5pt]
 \hline
 $\mathcal{F}^{(2)}_{F,h;\text{vac}}$ & 0 & 0 & $-\frac{1}{8}$ & $\frac{7}8  - \frac{\pi^2}{24}$ &  $-2.25015$\\[5pt]
 \hline
 $\mathcal{F}^{(2)}_{F,l;\text{vac}}$ & 0 & 0 & 0 & $\frac{17}{24}-\frac{\pi^2}{24}$ &  $-3.11684$ \\[5pt]
 \hline
 $\mathcal{F}^{(2)}_{A,h;\text{vac}}$ & 0 & 0 & 0 & 0 & 0 \\[5pt]
 \hline
 $\mathcal{F}^{(2)}_{A,l;\text{vac}}$ & 0 & 0 & 0 & 0 & 0 \\[5pt]
 \hline
\end{tabular}
\end{center}
\begin{equation*}
  \mathcal{F}^{(2,0)}_{\gamma \gamma, [1];\text{lbl}}=\left(0.64697 + i\, 2.07358\right) C_F T_F n_h+\left(0.73128 - i\, 1.79085\right) C_F T_F \tilde{n}_l
\end{equation*}

\vspace{0.7cm}
$gg \leftrightarrow {^1S_0^{[1]}}$:
\begin{center}
\renewcommand{\arraystretch}{1.5}
\begin{tabular}{| m{1.1cm} || m{0.7cm} | m{2cm} | m{2.5cm} | m{4.5cm} | m{1.5cm} |} 
 \hline
   $gg, [1]$ & $\epsilon^{-4}$ & $\epsilon^{-3}$ & $\epsilon^{-2}$ & $\epsilon^{-1}$ & $\epsilon^{0}$ \\[5pt]
 \hline\hline
 $\mathcal{F}^{(2)}_{FF}$ & 0 & 0 & $\frac{3}{32}$ & $-\frac{39}{32}-\frac{\pi^2}{16}+\frac{3}{4}\log{2}$ & $-9.58245$ \\[5pt]
 \hline
 $\mathcal{F}^{(2)}_{FA}$ & 0 & $-\frac{3}{16}$ & $\frac{7}{16}-\frac{\pi^2}{16}+\frac{3}{8}\log{2}$ & $-\frac{313}{96}+\frac{5i \pi}{4}-\frac{13\pi^2}{96}-\frac{i\pi^3}{16}-\frac{49}{16}\log{2}-\frac{3}{8}\log^2{2}+\frac{1}{8}\pi^2 \log{2}-\frac{7}{8}\zeta_3$ & $6.32284 - i\, 12.72196$\\[5pt]
 \hline
 $\mathcal{F}^{(2)}_{AA}$ & $\frac{1}{8}$ & $-\frac{11}{96}+\frac{i\pi}{4}-\frac{1}{2}\log{2}$ & $-\frac{139}{288}-\frac{11 i \pi}{48}-\frac{19\pi^2}{96}+\frac{11}{24}\log{2}-i\pi \log{2}+\log^2{2}$ & $-\frac{211}{432}-\frac{175 i \pi}{144}+\frac{7\pi^2}{64}-\frac{i\pi^3}{8}+\frac{175}{72}\log{2}+\frac{11}{12}i\pi \log{2}+\frac{11}{12}\pi^2 \log{2}-\frac{11}{12}\log^2{2} + 2i \pi \log^2{2}-\frac{4}{3} \log^3{2}+\frac{17}{48} \zeta_{3}$ & $3.45556 + i\, 24.90661$ \\[5pt]
 \hline
 $\mathcal{F}^{(2)}_{F,h;\text{vac}}$ & 0 & 0 & $-\frac{1}{8}$ & $\frac{7}8  - \frac{\pi^2}{24}$ & $-2.25015$ \\[5pt]
 \hline
 $\mathcal{F}^{(2)}_{F,l;\text{vac}}$ & 0 & 0 & 0 & $\frac{17}{24}-\frac{\pi^2}{24}$ & $-3.11684$ \\[5pt]
 \hline
 $\mathcal{F}^{(2)}_{A,h;\text{vac}}$ & 0 & $\frac{1}{6}$ & $\frac{1}{16}+\frac{i\pi}{6}-\frac{1}{3}\log{2}$ & $-\frac{7}{32}-\frac{\pi^2}{24}-\frac{1}{3}i\pi \log{2}+\frac{1}{3}\log^2{2}$ & $1.01362 - i\, 1.66960$ \\[5pt]
 \hline
 $\mathcal{F}^{(2)}_{A,l;\text{vac}}$ & 0 & $\frac{1}{24}$ & $\frac{5}{72}+\frac{i\pi}{12}-\frac{1}{6}\log{2}$ & $-\frac{17}{216}+\frac{5i\pi}{36}-\frac{\pi^2}{16}-\frac{5}{18}\log{2}-\frac{1}{3}i\pi \log{2}+\frac{1}{3}\log^2{2}$ & $1.41108 - i\, 4.80147$ \\[5pt]
 \hline
\end{tabular}
\end{center}
\begin{equation*}
\begin{split}
  \mathcal{F}^{(2,0)}_{gg, [1];\text{lbl}}&=\left(0.64697 + i\, 2.07358\right) C_F T_F n_h+\left(0.73128 - i\, 1.79085\right) C_F T_F \tilde{n}_l
  \\
  &+\left(0.17355 + i\, 0.27096\right) C_A T_F n_h+\left(-0.27562 + i\, 0.56535\right) C_A T_F \tilde{n}_l
\end{split}
\end{equation*}

\newpage
$\gamma g \leftrightarrow {^1S_0^{[8]}}$:
\begin{center}
\renewcommand{\arraystretch}{1.5}
\begin{tabular}{| m{1.1cm} || m{0.7cm} | m{2cm} | m{2.5cm} | m{4.5cm} | m{1.5cm} |} 
 \hline
  $\gamma g, [8]$ & $\epsilon^{-4}$ & $\epsilon^{-3}$ & $\epsilon^{-2}$ & $\epsilon^{-1}$ & $\epsilon^{0}$ \\[5pt]
 \hline\hline
 $\mathcal{F}^{(2)}_{FF}$ & 0 & 0 & $\frac{3}{32}$ & $-\frac{39}{32}-\frac{\pi^2}{16}+\frac{3}{4}\log{2}$ & $-9.58245$ \\[5pt]
 \hline
 $\mathcal{F}^{(2)}_{FA}$ & 0 & $-\frac{3}{16}$ & $\frac{21}{32}-\frac{\pi^2}{32}+\frac{3}{8}\log{2}$ & $-\frac{199}{96}+\frac{5\pi^2}{96}-\frac{77}{32}\log{2}+\frac{\pi^2}{16}\log{2}-\frac{3}{8}\log^2{2}-\frac{7}{16}\zeta_3$ & $11.58260$ \\[5pt]
 \hline
 $\mathcal{F}^{(2)}_{AA}$ & $\frac{1}{32}$ & $\frac{1}{192}-\frac{1}{8}\log{2}$ & $-\frac{223}{576}-\frac{1}{48}\log{2}+\frac{1}{4}\log^2{2}$ & $\frac{25}{108}+\frac{29\pi^2}{384}+\frac{187}{144}\log{2}+\frac{1}{24}\log^2{2}-\frac{1}{3}\log^3{2}-\frac{17}{96}\zeta_3$ & $-0.78451$ \\[5pt]
 \hline
 $\mathcal{F}^{(2)}_{F,h;\text{vac}}$ & 0 & 0 & $-\frac{1}{8}$ & $\frac{7}8  - \frac{\pi^2}{24}$ & $-2.25015$ \\[5pt]
 \hline
 $\mathcal{F}^{(2)}_{F,l;\text{vac}}$ & 0 & 0 & 0 & $\frac{17}{24}-\frac{\pi^2}{24}$ & $-3.11684$ \\[5pt]
 \hline
 $\mathcal{F}^{(2)}_{A,h;\text{vac}}$ & 0 & $\frac{1}{12}$ & $\frac{11}{96}-\frac{1}{6}\log{2}$ & $-\frac{53}{192}-\frac{1}{6}\log{2}+\frac{1}{6}\log^2{2}$ & $0.69685$ \\[5pt]
 \hline
 $\mathcal{F}^{(2)}_{A,l;\text{vac}}$ & 0 & $\frac{1}{48}$ & $\frac{11}{144}-\frac{1}{12}\log{2}$ & $-\frac{59}{432}-\frac{\pi^2}{96}-\frac{11}{36}\log{2}+\frac{1}{6}\log^2{2}$ & $0.15016$ \\[5pt]
 \hline
\end{tabular}
\end{center}
\begin{equation*}
\begin{split}
  \mathcal{F}^{(2,0)}_{\gamma g, [8];\text{lbl}}&=\left(1.29393 + i\, 4.14715\right) C_F T_F n_h+\left(1.46257 - i\, 3.58169\right) C_F T_F \tilde{n}_l
  \\
  &+\left(-0.48522 - i\, 1.55518\right) C_A T_F n_h+\left(-0.54846 + i\, 1.34313\right) C_A T_F \tilde{n}_l
\end{split}
\end{equation*}

\vspace{0.7cm}
$gg \leftrightarrow {^1S_0^{[8]}}$:
\begin{center}
\renewcommand{\arraystretch}{1.5}
\begin{tabular}{| m{1.1cm} || m{0.7cm} | m{2cm} | m{2.5cm} | m{4.5cm} | m{1.5cm} |} 
 \hline
   $gg, [8]$ & $\epsilon^{-4}$ & $\epsilon^{-3}$ & $\epsilon^{-2}$ & $\epsilon^{-1}$ & $\epsilon^{0}$ \\[5pt]
 \hline\hline
 $\mathcal{F}^{(2)}_{FF}$ & 0 & 0 & $\frac{3}{32}$ & $-\frac{39}{32}-\frac{\pi^2}{16}+\frac{3}{4}\log{2}$ & $-9.58245$ \\[5pt]
 \hline
 $\mathcal{F}^{(2)}_{FA}$ & 0 & $-\frac{9}{32}$ & $\frac{7}{8}-\frac{\pi^2}{16}+\frac{9}{16}\log{2}$ & $-\frac{253}{96}+\frac{5i\pi}{8}-\frac{\pi^2}{96}-\frac{i\pi^3}{32}-\frac{63}{16}\log{2}+\frac{1}{8}\pi^2\log{2}-\frac{9}{16}\log^2{2}-\frac{7}{8}\zeta_3$ & $13.9607 - i\, 6.3610$ \\[5pt]
 \hline
 $\mathcal{F}^{(2)}_{AA}$ & $\frac{1}{8}$ & $\frac{1}{96}+\frac{i\pi}{8}-\frac{1}{2}\log{2}$ & $-\frac{235}{288}-\frac{5i\pi}{96}-\frac{7\pi^2}{96}-\frac{1}{24}\log{2}-\frac{i\pi}{2}\log{2}+\log^2{2}$ & $\frac{43}{864}-\frac{211i \pi}{288}+\frac{29\pi^2}{192}-\frac{3i\pi^3}{64}+\frac{217}{72}\log{2}+\frac{5i\pi}{24}\log{2}+\frac{17\pi^2}{48}\log{2}+\frac{1}{12}\log^2{2}+i\pi\log^2{2}-\frac{4}{3}\log^3{2}-\frac{7}{48}\zeta_3$ & $-0.80634 + i\, 9.46843$ \\[5pt]
 \hline
 $\mathcal{F}^{(2)}_{F,h;\text{vac}}$ & 0 & 0 & $-\frac{1}{8}$ & $\frac{7}8  - \frac{\pi^2}{24}$ & $-2.25015$ \\[5pt]
 \hline
 $\mathcal{F}^{(2)}_{F,l;\text{vac}}$ & 0 & 0 & 0 & $\frac{17}{24}-\frac{\pi^2}{24}$ & $-3.11684$ \\[5pt]
 \hline
 $\mathcal{F}^{(2)}_{A,h;\text{vac}}$ & 0 & $\frac{1}{6}$ & $\frac{7}{48}+\frac{i\pi}{12}-\frac{1}{3}\log{2}$ & $-\frac{37}{96}-\frac{\pi^2}{48}-\frac{1}{6}\log{2}-\frac{i\pi}{6}\log{2}+\frac{1}{3}\log^2{2}$ & $1.20366 - i\, 0.83480$ \\[5pt]
 \hline
 $\mathcal{F}^{(2)}_{A,l;\text{vac}}$ & 0 & $\frac{1}{24}$ & $\frac{1}{9}+\frac{i\pi}{24}-\frac{1}{6}\log{2}$ & $-\frac{19}{108}+\frac{5i\pi}{72}-\frac{\pi^2}{24}-\frac{4}{9}\log{2}-\frac{i\pi}{6}\log{2}+\frac{1}{3}\log^2{2}$ & $0.85570 - i\, 2.40073$ \\[5pt]
 \hline
\end{tabular}
\end{center}
\begin{equation*}
\begin{split}
  \mathcal{F}^{(2,0)}_{gg, [8];\text{lbl}}&=\left(1.29393 + i\, 4.14715\right) C_F T_F n_h+\left(1.46257 - i\, 3.58169\right) C_F T_F \tilde{n}_l
  \\
  &+\left(-0.39845 - i\, 1.41970\right) C_A T_F n_h+\left(-0.68628 + i\, 1.62581\right) C_A T_F \tilde{n}_l
\end{split}
\end{equation*}

\newpage

\section{Renormalised amplitude structure}
\label{sec:renpole}

In this appendix, we give the structure of the renormalised amplitude for each form factor at one-loop level up to $\mathcal{O}{\left(\epsilon^2\right)}$ and at two-loop level up to the finite piece $\mathcal{O}{\left(\epsilon^0\right)}$. We proceed in a similar fashion as done in appendix~\ref{sec:barepole}. The decomposition follows the notation in  \cref{sec:UV}. From the finite piece on, we give for convenience the coefficients evaluated at the scale $\mu=m_Q$.

At one-loop order, we have the following results for the different form factors:

\begin{center}
\renewcommand{\arraystretch}{1.5}
\begin{tabular}{| m{1cm} || m{1cm} | m{2.0cm} | m{2.5cm} | m{4.0cm} | m{1.5cm} |} 
 \hline
 $\gamma \gamma, [1]$ & $\epsilon^{-2}$ & $ \epsilon^{-1}$ & $\left.\epsilon^{0}\right\vert_{\mu=m_Q}$ & $\left.\epsilon^{1}\right\vert_{\mu=m_Q}$ & $\left.\epsilon^{2}\right\vert_{\mu=m_Q}$ \\[5pt]
 \hline\hline
 $\overline{\mathcal{F}}^{(1)}_F$ & 0 & 0 & $\frac{\pi^2}{8}-\frac{5}{2}$ & $-1+\frac{1}{4}\pi^2+4 \log{2}+\frac{7}{4} \zeta_3$ & $5.52395$ \\[5pt]
 \hline
 $\overline{\mathcal{F}}^{(1)}_A$ & 0 & 0 & 0 & 0 & 0\\[5pt]
 \hline
 $\overline{\mathcal{F}}^{(1)}_l$ & 0 & 0 & 0 & 0 & 0 \\[5pt]
 \hline
\end{tabular}
\end{center}

\begin{center}
\renewcommand{\arraystretch}{1.5}
\begin{tabular}{| m{1cm} || m{1cm} | m{2.0cm} | m{2.5cm} | m{4.0cm} | m{1.5cm} |} 
 \hline
 $gg, [1]$ & $\epsilon^{-2}$ & $ \epsilon^{-1}$ & $\left.\epsilon^{0}\right\vert_{\mu=m_Q}$ & $\left.\epsilon^{1}\right\vert_{\mu=m_Q}$ & $\left.\epsilon^{2}\right\vert_{\mu=m_Q}$ \\[5pt]
 \hline\hline
 $\overline{\mathcal{F}}^{(1)}_F$ & 0 & 0 & $\frac{\pi^2}{8}-\frac{5}{2}$ & $-1+\frac{1}{4}\pi^2+4 \log{2}+\frac{7}{4} \zeta_3$ & $5.52395$ \\[5pt]
 \hline
 $\overline{\mathcal{F}}^{(1)}_A$ & $-\frac{1}{2}$ & $-\frac{i\pi}{2}+\log{2}-\frac{11}{12}-\frac{1}{2}l_{\mu}$ & $\frac{1}{2}+\frac{1}{6}\pi^2+i \pi \log{2}-\log^2{2}$ & $1+i \pi+\frac{1}{8}\pi^2+\frac{i}{8} \pi^3-2 \log{2}-\frac{7}{12} \pi^2 \log{2}-i \pi \log^2{2}+\frac{2}{3} \log^3{2}-\frac{7}{12} \zeta_3$ & $-9.59291 + i\, 4.79988$ \\[5pt]
 \hline
 $\overline{\mathcal{F}}^{(1)}_l$ & 0 & $\frac{1}{3}$ & 0 & 0 & 0 \\[5pt]
 \hline
\end{tabular}
\end{center}

\begin{center}
\renewcommand{\arraystretch}{1.5}
\begin{tabular}{| m{1cm} || m{1cm} | m{2.0cm} | m{2.5cm} | m{4.0cm} | m{1.5cm} |} 
 \hline
 $\gamma g, [8]$ & $\epsilon^{-2}$ & $ \epsilon^{-1}$ & $\left.\epsilon^{0}\right\vert_{\mu=m_Q}$ & $\left.\epsilon^{1}\right\vert_{\mu=m_Q}$ & $\left.\epsilon^{2}\right\vert_{\mu=m_Q}$ \\[5pt]
 \hline\hline
 $\overline{\mathcal{F}}^{(1)}_F$ & 0 & 0 & $\frac{\pi^2}{8}-\frac{5}{2}$ & $-1+\frac{1}{4}\pi^2+4 \log{2}+\frac{7}{4} \zeta_3$ & $5.52395$ \\[5pt]
 \hline
 $\overline{\mathcal{F}}^{(1)}_A$ & $-\frac{1}{4}$ & $-\frac{17}{24}+\frac{1}{2}\log{2}-\frac{1}{4}l_{\mu}$ & $\frac{3}{4}+ \frac{1}{48}\pi^2+\frac{1}{2}\log{2}-\frac{1}{2}\log^2{2}$ & $-\frac{1}{2}-\frac{1}{6}\pi^2-\frac{1}{2}\log{2}-\frac{1}{24} \pi^2 \log{2}-\frac{1}{2}\log^2{2}+\frac{1}{3}\log^3{2}+\frac{7}{12} \zeta_3$ & $-1.12737$ \\[5pt]
 \hline
 $\overline{\mathcal{F}}^{(1)}_l$ & 0 & $\frac{1}{6}$ & 0 & 0 & 0 \\[5pt]
 \hline
\end{tabular}
\end{center}

\begin{center}
\renewcommand{\arraystretch}{1.5}
\begin{tabular}{| m{1cm} || m{1cm} | m{2.0cm} | m{2.5cm} | m{4.0cm} | m{1.5cm} |} 
 \hline
 $gg, [8]$ & $\epsilon^{-2}$ & $ \epsilon^{-1}$ & $\left.\epsilon^{0}\right\vert_{\mu=m_Q}$ & $\left.\epsilon^{1}\right\vert_{\mu=m_Q}$ & $\left.\epsilon^{2}\right\vert_{\mu=m_Q}$ \\[5pt]
 \hline\hline
 $\overline{\mathcal{F}}^{(1)}_F$ & 0 & 0 & $\frac{\pi^2}{8}-\frac{5}{2}$ & $-1+\frac{1}{4}\pi^2+4 \log{2}+\frac{7}{4} \zeta_3$ & $5.52395$ \\[5pt]
 \hline
 $\overline{\mathcal{F}}^{(1)}_A$ & $-\frac{1}{2}$ & $-\frac{i\pi}{4}-\frac{7}{6}+\log{2}-\frac{1}{2}l_{\mu}$ & $1+\frac{5}{48} \pi^2+\frac{1}{2}\log{2}+\frac{1}{2} i \pi \log{2}-\log^2{2}$ & $\frac{i}{2} \pi-\frac{5}{48} \pi^2+\frac{i}{16} \pi^3-\frac{3}{2} \log{2}-\frac{1}{3} \pi^2 \log{2}-\frac{1}{2}\log^2{2}-\frac{1}{2} i \pi \log^2{2}+\frac{2}{3} \log^3{2}+\frac{7}{24} \zeta_3$ & $-5.92383+i\, 2.39994 $\\[5pt]
 \hline
 $\overline{\mathcal{F}}^{(1)}_l$ & 0 & $\frac{1}{3}$ & 0 & 0 & 0\\[5pt]
 \hline
\end{tabular}
\end{center}

\newpage

Similarly, at two-loop order we find the following structure for renormalised amplitudes. Since the light-by-light contributions are finite in four dimensions, we have that $\mathcal{\overline{F}}_{p,c;\text{lbl}}^{(2,0)}=\mathcal{F}_{p,c;\text{lbl}}^{(2,0)}$ and the corresponding values can be found in appendix~\ref{sec:barepole}.

\begin{center}
\renewcommand{\arraystretch}{1.5}
\begin{tabular}{| m{1.1cm} || m{0.7cm} | m{2cm} | m{2.5cm} | m{4.5cm} | m{1.5cm} |}  
 \hline
 $\gamma \gamma, [1]$  & $\epsilon^{-4}$ & $\epsilon^{-3}$ & $\epsilon^{-2}$ & $\epsilon^{-1}$ & $\left.\epsilon^{0}\right\vert_{\mu=m_Q}$ \\[5pt]
 \hline\hline
 $\overline{\mathcal{F}}^{(2)}_{FF}$ & 0 & 0 & 0 & $-\frac{\pi^2}{4}$ & $-21.10790$ \\[5pt]
 \hline
 $\overline{\mathcal{F}}^{(2)}_{FA}$ & 0 & 0 & 0 & $-\frac{\pi^2}{8}$ & $-4.79298$ \\[5pt]
 \hline
 $\overline{\mathcal{F}}^{(2)}_{AA}$ & 0 & 0 & 0 & 0 & 0 \\[5pt]
 \hline
 $\overline{\mathcal{F}}^{(2)}_{F,h;\text{vac}}$ & 0 & 0 & 0 & 0 & 0.22367 \\[5pt]
 \hline
 $\overline{\mathcal{F}}^{(2)}_{F,l;\text{vac}}$ & 0 & 0 & 0 & 0 & $-0.56482$ \\[5pt]
 \hline
 $\overline{\mathcal{F}}^{(2)}_{A,h;\text{vac}}$ & 0 & 0 & 0 & 0 & 0 \\[5pt]
 \hline
 $\overline{\mathcal{F}}^{(2)}_{A,l;\text{vac}}$ & 0 & 0 & 0 & 0 & 0 \\[5pt]
 \hline
 $\overline{\mathcal{F}}^{(2)}_{l,l;\text{vac}}$ & 0 & 0 & 0 & 0 & 0 \\[5pt]
 \hline
\end{tabular}
\end{center}

\begin{center}
\renewcommand{\arraystretch}{1.5}
\begin{tabular}{| m{1.1cm} || m{0.7cm} | m{2cm} | m{2.5cm} | m{4.5cm} | m{1.5cm} |} 
 \hline
 $gg, [1]$ & $\epsilon^{-4}$ & $\epsilon^{-3}$ & $\epsilon^{-2}$ & $\epsilon^{-1}$ & $\left.\epsilon^{0}\right\vert_{\mu=m_Q}$ \\[5pt]
 \hline\hline
 $\overline{\mathcal{F}}^{(2)}_{FF}$ & 0 & 0 & 0 & $-\frac{\pi^2}{4}$ & $-21.10790$ \\[5pt]
 \hline
 $\overline{\mathcal{F}}^{(2)}_{FA}$ & 0 & 0 & $\frac{5}{4}-\frac{\pi^2}{16}$ & $\frac{67}{24}+\frac{5i \pi}{4}-\frac{35\pi^2}{96}-\frac{i\pi^3}{16}-\frac{9}{2}\log{2}+\frac{1}{8}\pi^2 \log{2}+\frac{5}{2}l_{\mu}-\frac{\pi^2}{8}l_{\mu}-\frac{7}{8}\zeta_3$ & $-5.81386 - i\, 12.72196$ \\[5pt]
 \hline
 $\overline{\mathcal{F}}^{(2)}_{AA}$ & $\frac{1}{8}$ & $\frac{77}{96}+\frac{i\pi}{4}-\frac{1}{2}\log{2}+\frac{1}{4}l_{\mu}$ & $\frac{103}{288}+\frac{11 i \pi}{16}-\frac{19\pi^2}{96}-\frac{11}{8}\log{2}-i\pi \log{2}+\log^2{2}+\frac{11}{16}l_{\mu}+\frac{i\pi}{2}l_{\mu}-\log{2}\, l_{\mu}+\frac{1}{4}l_{\mu}^2$ & $-\frac{95}{54}-\frac{175 i \pi}{144}-\frac{113\pi^2}{576}-\frac{i\pi^3}{8}+\frac{175}{72}\log{2}-\frac{11}{12}i\pi \log{2}+\frac{11}{12}\pi^2 \log{2}+\frac{11}{12}\log^2{2}+2i \pi \log^2{2}-\frac{4}{3}\log^3{2}-\frac{139}{144}l_{\mu}+\frac{11}{24}i\pi l_{\mu}-\frac{19}{48}\pi^2 l_{\mu}-\frac{11}{12}\log{2}l_{\mu}-2i\pi\log{2}l_{\mu}+2\log^2{2}l_{\mu}+\frac{11}{48}l^2_{\mu}+\frac{i\pi}{2}l_{\mu}^2-\log{2}l_{\mu}^2+\frac{1}{6}l_{\mu}^3+\frac{17}{48}\zeta_3$ & $10.09665 + i\, 14.80863$ \\[5pt]
 \hline
 $\overline{\mathcal{F}}^{(2)}_{F,h;\text{vac}}$ & 0 & 0 & 0 & 0 & 0.22367 \\[5pt]
 \hline
 $\overline{\mathcal{F}}^{(2)}_{F,l;\text{vac}}$ & 0 & 0 & 0 & $-\frac{17}{24}+\frac{\pi^2}{24}$ & 1.54971 \\[5pt]
 \hline
 $\overline{\mathcal{F}}^{(2)}_{A,h;\text{vac}}$ & 0 & 0 & 0 & 0 & $-0.19436 - i\, 0.26425$ \\[5pt]
 \hline
 $\overline{\mathcal{F}}^{(2)}_{A,l;\text{vac}}$ & 0 & $-\frac{7}{24}$ & $-\frac{13}{24}-\frac{i\pi}{4}+\frac{1}{2}\log{2}-\frac{1}{4}l_{\mu}$ & $\frac{25}{54}+\frac{5i\pi}{36}+\frac{7\pi^2}{144}-\frac{5}{18}\log{2}+\frac{1}{3}i\pi \log{2}-\frac{1}{3}\log^2{2}+\frac{5}{36}l_{\mu}-\frac{i\pi}{6}l_{\mu}+\frac{1}{3}\log{2}l_{\mu}-\frac{1}{12}l_{\mu}^2$ & $-1.00386 - i\, 1.12947$\\[5pt]
 \hline
 $\overline{\mathcal{F}}^{(2)}_{l,l;\text{vac}}$ & 0 & 0 & $\frac{1}{9}$ & 0 & 0 \\[5pt]
 \hline
\end{tabular}
\end{center}

\newpage

\begin{center}
\renewcommand{\arraystretch}{1.5}
\begin{tabular}{| m{1.1cm} || m{0.7cm} | m{2cm} | m{2.5cm} | m{4.5cm} | m{1.5cm} |} 
 \hline
 $\gamma g, [8]$ & $\epsilon^{-4}$ & $\epsilon^{-3}$ & $\epsilon^{-2}$ & $\epsilon^{-1}$ & $\left.\epsilon^{0}\right\vert_{\mu=m_Q}$ \\[5pt]
 \hline\hline
 $\overline{\mathcal{F}}^{(2)}_{FF}$ & 0 & 0 & 0 & $-\frac{\pi^2}{4}$ & $-21.10790$ \\[5pt]
 \hline
 $\overline{\mathcal{F}}^{(2)}_{FA}$ & 0 & 0 & $\frac{5}{8}-\frac{\pi^2}{32}$ & $\frac{97}{48}-\frac{5\pi^2}{192}-\frac{9}{4}\log{2}+\frac{\pi^2}{16}\log{2}+\frac{5}{4}l_{\mu}-\frac{\pi^2}{16}l_{\mu}-\frac{7}{16}\zeta_3$ & $7.82058$ \\[5pt]
 \hline
 $\overline{\mathcal{F}}^{(2)}_{AA}$ & $\frac{1}{32}$ & $\frac{67}{192}-\frac{1}{8}\log{2}+\frac{1}{16}l_{\mu}$ & $\frac{313}{1152}-\frac{17}{24}\log{2}+\frac{1}{4}\log^2{2}+\frac{17}{48}l_{\mu}-\frac{1}{4}\log{2}l_{\mu}+\frac{1}{16}l_{\mu}^2$ & $-\frac{211}{216}+\frac{3\pi^2}{64}+\frac{11}{18}\log{2}+\frac{35}{48}\log^2{2}-\frac{1}{3}\log^3{2}-\frac{31}{72}l_{\mu}-\frac{35}{48}\log{2}l_{\mu}+\frac{1}{2}\log^2{2}l_{\mu}+\frac{35}{192}l_{\mu}^2-\frac{1}{4}\log{2}l_{\mu}^2+\frac{1}{24}l_{\mu}^3-\frac{17}{96}\zeta_3$ & $2.24678$ \\[5pt]
 \hline
 $\overline{\mathcal{F}}^{(2)}_{F,h;\text{vac}}$ & 0 & 0 & 0 & 0 & $0.22367$ \\[5pt]
 \hline
 $\overline{\mathcal{F}}^{(2)}_{F,l;\text{vac}}$ & 0 & 0 & 0 & $-\frac{17}{48}+\frac{\pi^2}{48}$ & $0.49245$ \\[5pt]
 \hline
 $\overline{\mathcal{F}}^{(2)}_{A,h;\text{vac}}$ & 0 & 0 & 0 & 0 & $-0.10680$ \\[5pt]
 \hline
 $\overline{\mathcal{F}}^{(2)}_{A,l;\text{vac}}$ & 0 & $-\frac{5}{48}$ & $-\frac{5}{18}+\frac{1}{6}\log{2}-\frac{1}{12}l_{\mu}$ & $\frac{37}{108}-\frac{1}{18}\log{2}-\frac{1}{12}\log^2{2}+\frac{1}{36}l_{\mu}+\frac{1}{12}\log{2} l_{\mu}-\frac{1}{48}l_{\mu}^2$ & $-0.95213$ \\[5pt]
 \hline
 $\overline{\mathcal{F}}^{(2)}_{l,l;\text{vac}}$ & 0 & 0 & $\frac{1}{24}$ & 0 & 0 \\[5pt]
 \hline
\end{tabular}
\end{center}

\begin{center}
\renewcommand{\arraystretch}{1.5}
\begin{tabular}{| m{1.1cm} || m{0.7cm} | m{2cm} | m{2.5cm} | m{4.5cm} | m{1.5cm} |} 
 \hline
 $gg, [8]$ & $\epsilon^{-4}$ & $\epsilon^{-3}$ & $\epsilon^{-2}$ & $\epsilon^{-1}$ & $\left.\epsilon^{0}\right\vert_{\mu=m_Q}$ \\[5pt]
 \hline\hline
 $\overline{\mathcal{F}}^{(2)}_{FF}$ & 0 & 0 & 0 & $-\frac{\pi^2}{4}$ & $-21.10790$ \\[5pt]
 \hline
 $\overline{\mathcal{F}}^{(2)}_{FA}$ & 0 & 0 & $\frac{5}{4}-\frac{\pi^2}{16}$ & $\frac{41}{12}+\frac{5i\pi}{8}-\frac{7\pi^2}{48}-\frac{i\pi^3}{32}-\frac{9}{2}\log{2}+\frac{1}{8}\pi^2\log{2}+\frac{5}{2}l_{\mu}-\frac{1}{8}\pi^2 l_{\mu}-\frac{7}{8}\zeta_3$ & $7.31014 - i\, 6.36098$ \\[5pt]
 \hline
 $\overline{\mathcal{F}}^{(2)}_{AA}$ & $\frac{1}{8}$ & $\frac{89}{96}+\frac{i\pi}{8}-\frac{1}{2}\log{2}+\frac{1}{4}l_{\mu}$ & $\frac{139}{288}+\frac{13i\pi}{32}-\frac{7\pi^2}{96}-\frac{15}{8}\log{2}-\frac{i\pi}{2}\log{2}+\log^2{2}+\frac{15}{16}l_{\mu}+\frac{i\pi}{4}l_{\mu}-\log{2}\, l_{\mu}+\frac{1}{4}l_{\mu}^2$ & $-\frac{1847}{864}-\frac{211i \pi}{288}-\frac{23\pi^2}{576}-\frac{3i\pi^3}{64}+\frac{151}{72}\log{2}-\frac{17i\pi}{24}\log{2}+\frac{17\pi^2}{48}\log{2}+\frac{23}{12}\log^2{2}+i\pi\log^2{2}-\frac{4}{3}\log^3{2}-\frac{169}{144}l_{\mu}+\frac{17i\pi}{48}l_{\mu}-\frac{7}{48}\pi^2 l_{\mu}-\frac{23}{12}\log{2}l_{\mu}-i\pi \log{2} l_{\mu}+2\log^2{2}l_{\mu}+\frac{23}{48}l^2_{\mu}+\frac{i\pi}{4}l^2_{\mu}-\log{2}l^2_{\mu}+\frac{1}{6}l_{\mu}^3-\frac{7}{48}\zeta_3$ & $6.55592 + i\, 4.41944$ \\[5pt]
 \hline
 $\overline{\mathcal{F}}^{(2)}_{F,h;\text{vac}}$ & 0 & 0 & 0 & 0 & $0.22367$ \\[5pt]
 \hline
 $\overline{\mathcal{F}}^{(2)}_{F,l;\text{vac}}$ & 0 & 0 & 0 & $-\frac{17}{24}+\frac{\pi^2}{24}$ & $1.54971$ \\[5pt]
 \hline
 $\overline{\mathcal{F}}^{(2)}_{A,h;\text{vac}}$ & 0 & 0 & 0 & 0 & $-0.20398 - i\, 0.13212$ \\[5pt]
 \hline
 $\overline{\mathcal{F}}^{(2)}_{A,l;\text{vac}}$ & 0 & $-\frac{7}{24}$ & $-\frac{2}{3}-\frac{i\pi}{8}+\frac{1}{2}\log{2}-\frac{1}{4}l_{\mu}$ & $\frac{151}{216}+\frac{5i\pi}{72}+\frac{\pi^2}{36}-\frac{1}{9}\log{2}+\frac{i\pi}{6}\log{2}-\frac{1}{3}\log^2{2}+\frac{1}{18}l_{\mu}-\frac{i\pi}{12}l_{\mu}+\frac{1}{3}\log{2}l_{\mu}-\frac{1}{12}l_{\mu}^2$ & $-1.82149 - i\, 0.56474$ \\[5pt]
 \hline
 $\overline{\mathcal{F}}^{(2)}_{l,l;\text{vac}}$ & 0 & 0 & $\frac{1}{9}$ & 0 & 0 \\[5pt]
 \hline
\end{tabular}
\end{center}

\section{Renormalisation coefficients}
\label{sec:RenCoef}

In this appendix, we give the expressions for the renormalisation coefficients $Z$ used to remove the UV singularities. We apply for the gluon wavefunction $Z_g$, the heavy-quark wavefunction $Z_Q$ and for the renormalisation of the heavy-quark mass $Z_m$ the on-shell renormalisation scheme, while for the coupling renormalisation $Z_{\alpha_s}$ we adopt the \msbar-scheme \cite{Barnreuther:2013qvf,Broadhurst:1993mw,Melnikov:2000qh,Mitov:2006xs}. In the following, the expansion in the strong coupling involves the coupling with $n_f=n_l+n_h$ flavours. In order to get to the conventional coupling where only $n_l$ massless flavours are absorbed, we need to apply the decoupling relation given in ref.~\cite{Bernreuther:1981sg}, which reads:
\begin{equation}
    \alpha_s^{\left(n_f\right)}=\zeta_{\alpha_{s}}\alpha_s^{\left(n_l\right)}\,,
\end{equation}
with
\begin{align}
    \zeta_{\alpha_s}=&1+\left(\frac{\alpha_s^{\left(n_l\right)}}{\pi}\right)T_F n_h \left[\frac{1}{3} l_{\mu}+\frac{1}{6}\epsilon l_{\mu}^2 + \frac{\pi^2}{36}\epsilon+\frac{1}{18}\epsilon^2 l_{\mu}^3 +\frac{\pi^2}{36}\epsilon^2 l_{\mu}-\frac{1}{9}\epsilon^2 \zeta_3\right] \nonumber
    \\
    &\hspace*{-0.3cm} +\left(\frac{\alpha_s^{\left(n_l\right)}}{\pi}\right)^2 T_F n_h \left[\frac{1}{9}T_F n_h l_{\mu}^2+C_F\left(\frac{15}{16}+\frac{1}{4}l_{\mu}\right)+C_A\left(-\frac{2}{9}+\frac{5}{12}l_{\mu}\right)\right] + \mathcal{O}{\left(\alpha_s^3\right)}\,.
\end{align}
The renormalisation factors read
\begin{align}
    Z_g=&1+\left(\frac{\alpha_s^{\left(n_f\right)}}{\pi}\right)T_F n_h \left[-\frac{1}{3\epsilon}-\frac{1}{3}l_{\mu}-\frac{1}{6}\epsilon l_{\mu}^2 - \frac{\pi^2}{36}\epsilon-\frac{1}{18}\epsilon^2 l_{\mu}^3-\frac{\pi^2}{36} \epsilon^2 l_{\mu}+\frac{1}{9}\epsilon^2 \zeta_3\right] \nonumber
    \\
    &\hspace*{-0.3cm} +\left(\frac{\alpha_s^{\left(n_f\right)}}{\pi}\right)^2 T_F n_h \left[T_F n_h\left(\frac{1}{9\epsilon}l_{\mu}+\frac{1}{6}l_{\mu}^2+\frac{\pi^2}{108}\right)+T_F n_l \left(-\frac{1}{9\epsilon^2}-\frac{1}{9\epsilon}l_{\mu}-\frac{1}{18}l_{\mu}^2-\frac{\pi^2}{108}\right)\right. \nonumber
    \\
    &\hspace*{-0.3cm} \left.+C_F\left(-\frac{1}{8\epsilon}-\frac{1}{4}l_{\mu}-\frac{15}{16}\right)+C_A\left(\frac{35}{144\epsilon^2}+\frac{13}{72}l_{\mu}-\frac{5}{32\epsilon}-\frac{5}{16}l_{\mu}+\frac{1}{36}l_{\mu}^2+\frac{13}{192}+\frac{13\pi^2}{864}\right)\right] \nonumber
    \\
    &\hspace*{-0.3cm} + \mathcal{O}{\left(\alpha_s^3\right)}\,,
\\[10pt]
    Z_Q=&1+\left(\frac{\alpha_s^{\left(n_f\right)}}{\pi}\right)C_F\left[-\frac{3}{4\epsilon}-1-\frac{3}{4}l_{\mu}-2\epsilon-\epsilon l_{\mu} -\frac{3}{8}\epsilon l_{\mu}^2-\frac{\pi^2}{16}\epsilon-4\epsilon^2-2\epsilon^2 l_{\mu}-\frac{1}{2}\epsilon^2 l_{\mu}^2-\frac{1}{8}\epsilon^2 l_{\mu}^3\right. \nonumber
    \\
    &\hspace*{-0.3cm} \left.-\frac{\pi^2}{12}\epsilon^2-\frac{\pi^2}{16}\epsilon^2 l_{\mu}+\frac{1}{4}\epsilon^2 \zeta_3\right]+\left(\frac{\alpha_s^{\left(n_f\right)}}{\pi}\right)^2 C_F \left[T_F n_h \left(\frac{1}{16\epsilon}+\frac{1}{4\epsilon}l_{\mu}+\frac{947}{288}+\frac{11}{24}l_{\mu}+\frac{3}{8}l_{\mu}^2-\frac{5\pi^2}{16}\right)\right. \nonumber
    \\
    &\hspace*{-0.3cm} \left.+T_F n_l \left(-\frac{1}{8\epsilon^2}+\frac{11}{48\epsilon}+\frac{113}{96}+\frac{19}{24}l_{\mu}+\frac{1}{8}l_{\mu}^2+\frac{\pi^2}{12}\right)+C_F\left(\frac{9}{32\epsilon^2}+\frac{51}{64\epsilon}+\frac{9}{16\epsilon}l_{\mu}+\frac{433}{128}\right.\right. \nonumber
    \\
    &\hspace*{-0.3cm} \left.\left.+\frac{51}{32}l_{\mu}+\frac{9}{16}l_{\mu}^2-\frac{49\pi^2}{64}+\pi^2\log{2}-\frac{3}{2}\zeta_3\right)+C_A\left(\frac{11}{32\epsilon^2}-\frac{127}{192\epsilon}-\frac{1705}{384}-\frac{215}{96}l_{\mu}-\frac{11}{32}l_{\mu}^2\right.\right. \nonumber
    \\
    &\hspace*{-0.3cm} \left.\left.+\frac{5\pi^2}{16}-\frac{1}{2}\pi^2 \log{2}+\frac{3}{4}\zeta_3\right)\right]+ \mathcal{O}{\left(\alpha_s^3\right)}\,,
\\[10pt]
    Z_m=&1+\left(\frac{\alpha_s^{\left(n_f\right)}}{\pi}\right)C_F\left[-\frac{3}{4\epsilon}-1-\frac{3}{4}l_{\mu}-2\epsilon-\epsilon l_{\mu}-\frac{3}{8}\epsilon l_{\mu}^2-\frac{\pi^2}{16}\epsilon-4\epsilon^2-2\epsilon^2 l_{\mu}-\frac{1}{2}\epsilon^2 l_{\mu}^2-\frac{1}{8}\epsilon^2 l_{\mu}^3\right. \nonumber
    \\
    &\hspace*{-0.3cm}\left.-\frac{\pi^2}{12}\epsilon^2-\frac{\pi^2}{16}\epsilon^2 l_{\mu}+\frac{1}{4}\epsilon^2 \zeta_3\right]+\left(\frac{\alpha_s^{\left(n_f\right)}}{\pi}\right)^2 C_F\left[T_F n_h \left(-\frac{1}{8\epsilon^2}+\frac{5}{48\epsilon}+\frac{143}{96}+\frac{13}{24}l_{\mu}+\frac{1}{8}l_{\mu}^2-\frac{\pi^2}{6}\right)\right. \nonumber
    \\
    &\hspace*{-0.3cm} +T_F n_l \left(-\frac{1}{8\epsilon^2}+\frac{5}{48\epsilon}+\frac{71}{96}+\frac{13}{24}l_{\mu}+\frac{1}{8}l_{\mu}^2+\frac{\pi^2}{12}\right)+C_F \left(\frac{9}{32\epsilon^2}+\frac{45}{64\epsilon}+\frac{9}{16\epsilon}l_{\mu}+\frac{199}{128}\right. \nonumber
    \\
    &\hspace*{-0.3cm} \left.+\frac{45}{32}l_{\mu}+\frac{9}{16}l_{\mu}^2-\frac{17\pi^2}{64}+\frac{1}{2}\pi^2 \log{2}-\frac{3}{4}\zeta_3\right)+C_A \left(\frac{11}{32\epsilon^2}-\frac{97}{192\epsilon}-\frac{1111}{384}-\frac{185}{96}l_{\mu}-\frac{11}{32}l_{\mu}^2\right. \nonumber
    \\
    &\hspace*{-0.3cm} \left.\left.+\frac{\pi^2}{12}-\frac{1}{4}\pi^2 \log{2}+\frac{3}{8}\zeta_3\right)\right]+ \mathcal{O}{\left(\alpha_s^3\right)}\,,
\\[10pt]
    Z_{\alpha_s}=&1-\left(\frac{\alpha_s^{\left(n_f\right)}}{\pi}\right)\frac{\beta_0}{4\epsilon}+\left(\frac{\alpha_s^{\left(n_f\right)}}{\pi}\right)^2 \left(\frac{\beta_0^2}{16\epsilon^2}-\frac{\beta_1}{32\epsilon}\right)+ \mathcal{O}{\left(\alpha_s^3\right)}\,,
\end{align}
with
\begin{equation}
    \beta_0=\frac{11}{3}C_A-\frac{4}{3}T_F n_f, \hspace{1cm} \beta_1=\frac{34}{3}C_A^2-\frac{20}{3}C_A T_F n_f-4 C_F T_F n_f.
\end{equation}

\section{IR singularities}
\label{sec:IRRenCoef}

In this appendix, we give the coefficients that appear in the anomalous dimension matrix $\mathbf{\Gamma}$ which is needed to construct the IR singularity structure $\mathbf{Z}_{\text{IR}}$ for the amplitude. The expansion in the coupling is done with $n_l$ light flavours inside the running. Apart from the coefficient $ \gamma_{\text{cusp}}^{\textrm{thres}}$, which we have computed in the main text, the remaining coefficients have been computed in refs.~\cite{Becher:2009qa,Korchemsky:1987wg,Korchemsky:1988hd,Korchemskaya:1992je,Kodaira:1981nh,Catani:1988vd,Becher:2009kw}. The coefficients read 
\begin{align}
    \gamma_{\text{cusp}}=&\left(\frac{\alpha_s^{\left(n_l\right)}}{\pi}\right)+\left(\frac{\alpha_s^{\left(n_l\right)}}{\pi}\right)^2\left[C_A \left(\frac{67}{36}-\frac{\pi^2}{12}\right)-\frac{5}{9}T_F n_l\right] + \mathcal{O}{\left(\alpha_s^3\right)}\,,
    \\
    \gamma^g=&-\left(\frac{\alpha_s^{\left(n_l\right)}}{\pi}\right)\frac{\beta_0}{4}+\left(\frac{\alpha_s^{\left(n_l\right)}}{\pi}\right)^2\left[C_A^2\left(-\frac{173}{108}+\frac{11\pi^2}{288}+\frac{1}{8}\zeta_3\right)\right. \nonumber
    \\
    &\left.+C_A T_F n_l \left(\frac{16}{27}-\frac{\pi^2}{72}\right)+\frac{1}{4}C_F T_F n_l\right]+ \mathcal{O}{\left(\alpha_s^3\right)}\,,
    \\
    \gamma^Q=&-\left(\frac{\alpha_s^{\left(n_l\right)}}{\pi}\right)\frac{C_F}{2}+\left(\frac{\alpha_s^{\left(n_l\right)}}{\pi}\right)^2\frac{C_F}{4}\left[C_A \left(-\frac{49}{18}+\frac{\pi^2}{6}-\zeta_3\right)+\frac{10}{9}T_F n_l\right]+ \mathcal{O}{\left(\alpha_s^3\right)}\,.
\end{align}

\section{Form factors: analytic expressions}
\label{sec:formfactoranalyticexpression}

In this appendix, we collect the analytical expressions for the individual coefficients in the form-factor decomposition defined in section~\ref{sec:formfac}. We also make them available in electronic form in ref.~\cite{formFactorsGit}. These coefficients can be expressed in terms of master integrals that we have computed in our companion paper \cite{Abreu:2022vei}. In the case where these are expressible in terms of multiple polylogarithms, we will write them out explicitly. For the integrals that involve functions in the class of elliptic multiple polylogarithms and iterated integrals of modular forms, as these are rather lengthy, we will keep the master integral notation. The master integrals will be expanded in the dimensional regulator $\epsilon$ as done in our companion paper \cite{Abreu:2022vei}
\begin{equation}
    F_I=\sum_{k} \epsilon^k F_I^{(k)}.
\end{equation}
The complete analytical expressions for the $F_I^{(k)}$ terms can be found in ref.~\cite{masterIntegralsGit}.

In the following we define some non-trivial constants that appear in the coefficients and that have not yet been defined previously in our companion paper \cite{Abreu:2022vei}. The Catalan constant $C$ is defined as
\begin{equation}
    C=\sum_{n=0}^{\infty}\frac{\left(-1\right)^n}{\left(2n+1\right)^2}=0.915966\ldots,
\end{equation}
whereas the polygamma function $\psi^m{\left(z\right)}$ is given by
\begin{equation}
    \psi^m{\left(z\right)}=\frac{d^{m+1}}{dz^{m+1}}\log{\left(\Gamma{\left(z\right)}\right)}
\end{equation}
with $\Gamma{\left(z\right)}$ being the gamma function. The function $\hpli{\left(0,-,+,-\right)}$ can be expressed in terms of multiple polylogarithms as
\begin{equation}
\begin{split}
    \hpli{\left(0,-,+,-\right)}=&G{\left(0,-1,-1,-1;i\right)}+G{\left(0,-1,-1,1;i\right)}-G{\left(0,-1,1,-1;i\right)}
    \\
    &-G{\left(0,-1,1,1;i\right)}+G{\left(0,1,-1,-1;i\right)}+G{\left(0,1,-1,1;i\right)}
    \\
    &-G{\left(0,1,1,-1;i\right)}-G{\left(0,1,1,1;i\right)}.
\end{split}
\end{equation}
Having defined the constants above, we now turn to the individual form-factor coefficients.

We first collect the coefficients $a_i^{(2)}$ needed for the regular contributions,
\begin{align}
	a_{1}^{(2)}=&\frac{1261}{96}-\frac{2579}{1728}\pi^2-\frac{57743}{4147200}\pi^4+\frac{35}{288}\pi\Cl_2{\left(\frac{\pi}{3}\right)}+\frac{3}{32}\pi\Ime{\left[G{\left(0,1,e^{-\frac{2i\pi}{3}};1\right)}\right]}\nonumber
    \\
    & \hspace{-1cm} +\frac{1}{8}\pi \Ime{\left[G{\left(0,e^{-\frac{i\pi}{3}},-1;1\right)}\right]}+\frac{7}{16}\Ree{\left[G{\left(e^{-\frac{i\pi}{3}},1,-1;1\right)}\right]}-\frac{681}{320}\Ree{\left[G{\left(0,0,e^{-\frac{i\pi}{3}},-1;1\right)}\right]}\nonumber
    \\
    & \hspace{-1cm} -\frac{2043}{1280}\Ree{\left[G{\left(0,0,e^{-\frac{2i\pi}{3}},1;1\right)}\right]}-\frac{3}{4}\Ree{\left[G{\left(e^{-\frac{i\pi}{3}},1,1,-1;1\right)}\right]}+\frac{11}{6}\log{2}-\frac{1253}{960}\pi^2 \log{2}
    \nonumber\\
    & \hspace{-1cm} +\frac{5}{48}\pi\Cl_2{\left(\frac{\pi}{3}\right)}\log{2}-\frac{3}{8}\Ree{\left[G{\left(e^{-\frac{i\pi}{3}},1,-1;1\right)}\right]}\log{2}-\frac{37}{24}\log^2{2}-\frac{43}{160}\pi^2\log^2{2}+\frac{7}{64}\log^3{2}
    \nonumber\\
    & \hspace{-1cm} -\frac{7}{32}\pi^2 \log{3}+\frac{1}{32}\pi \Cl_2{\left(\frac{\pi}{3}\right)}\log{3}-\frac{7}{32}\log^2{2}\,\log{3}-\frac{3}{16}\pi^2\Li_2{\left(-\frac{1}{2}\right)}+\frac{7}{32}\log{2}\,\Li_2{\left(-\frac{1}{2}\right)}
    \nonumber\\
    & \hspace{-1cm} +\frac{3923}{2880}\zeta_3+\frac{355}{96}\zeta_3\log{2}-F_3^{(0)}-3 F_{14}^{(0)}+\frac{7}{20} F_{15}^{(0)} -\frac{11}{6} F_{18}^{(0)}-F_{22}^{(0)}+\frac{31}{20} F_{31}^{(0)}+\frac{59}{15} F_{57}^{(0)}
    \nonumber\\
    & \hspace{-1cm} -\frac{29}{36} F_{64}^{(0)}+\frac{37}{36} F_{65}^{(0)},
    \\[12pt]
    a_{2}^{(2)}=& -\frac{4753}{576}+\frac{3181}{1728}\pi^2+\frac{159091}{6220800}\pi^4 -\frac{85}{576}\pi \Cl_{2}{\left(\frac{\pi}{3}\right)}-\frac{1}{8}\pi \Ime{\left[G{\left(0,1,e^{-\frac{2i\pi}{3}};1\right)}\right]} \nonumber
    \\
    & \hspace{-1cm} -\frac{1}{6}\pi \Ime{\left[G{\left(0,e^{-\frac{i\pi}{3}},-1;1\right)}\right]}-\frac{17}{32} \Ree{\left[G{\left(e^{-\frac{i\pi}{3}},1,-1;1\right)}\right]}+\frac{499}{160}\Ree{\left[G{\left(0,0,e^{-\frac{i\pi}{3}},-1;1\right)}\right]} \nonumber
    \\
    & \hspace{-1cm} +\frac{1497}{640} \Ree{\left[G{\left(0,0,e^{-\frac{2i\pi}{3}},1;1\right)}\right]} + \Ree{\left[G{\left(e^{-\frac{i\pi}{3}},1,1,-1;1\right)}\right]} + \frac{55}{24}\log{2}-\frac{1559}{640}\pi^2 \log{2} \nonumber
    \\
    & \hspace{-1cm} -\frac{5}{36}\pi \Cl_{2}{\left(\frac{\pi}{3}\right)} \log{2} +\frac{1}{2}\log{2}\Ree{\left[G{\left(e^{-\frac{i\pi}{3}},1,-1;1\right)}\right]}+\frac{11}{16}\log^2{2}+\frac{17}{80}\pi^2 \log^2{2} \nonumber
    \\
    & \hspace{-1cm} -\frac{17}{128}\log^3{2}+\frac{17}{64}\pi^2 \log{3}-\frac{1}{24}\log{3}\pi \Cl_{2}{\left(\frac{\pi}{3}\right)}+\frac{17}{64}\log^2{2} \log{3} +\frac{1}{4}\pi^2 \Li_2{\left(-\frac{1}{2}\right)} \nonumber
    \\
    & \hspace{-1cm} -\frac{17}{64} \Li_2{\left(-\frac{1}{2}\right)} \log{2} -\frac{45253}{5760}\zeta_3 -\frac{185}{144}\zeta_3 \log{2} +\frac{1}{2}F_3^{(0)}+\frac{3}{2}F_{14}^{(0)}-\frac{7}{40}F_{15}^{(0)}+\frac{11}{12}F_{18}^{(0)} \nonumber
    \\
    & \hspace{-1cm} +\frac{1}{2}F_{22}^{(0)}-\frac{31}{40}F_{31}^{(0)}-\frac{59}{30}F_{57}^{(0)}+\frac{29}{72}F_{64}^{(0)}-\frac{37}{72}F_{65}^{(0)},
    \\[12pt]
    a_{3}^{(2)}=& -\frac{10675}{576}+\frac{4225}{1728}\pi^2+\frac{391037}{6912000}\pi^4-\frac{275}{576}\pi \Cl_{2}{\left(\frac{\pi}{3}\right)}-\frac{9}{32}\pi \Ime{\left[G{\left(0,1,e^{-\frac{2i\pi}{3}};1\right)}\right]} \nonumber
    \\
    & \hspace{-1cm} -\frac{3}{8}\pi \Ime{\left[G{\left(0,e^{-\frac{i\pi}{3}},-1;1\right)}\right]} -\frac{55}{32} \Ree{\left[G{\left(e^{-\frac{i\pi}{3}},1,-1;1\right)}\right]}+\frac{15297}{1600}\Ree{\left[G{\left(0,0,e^{-\frac{i\pi}{3}},-1;1\right)}\right]} \nonumber
    \\
    & \hspace{-1cm} +\frac{45891}{6400}\Ree{\left[G{\left(0,0,e^{-\frac{2i\pi}{3}},-1;1\right)}\right]}+\frac{9}{4}\Ree{\left[G{\left(e^{-\frac{i\pi}{3}},1,1,-1;1\right)}\right]} +\frac{55}{24}\log{2}-\frac{5707}{1920}\pi^2 \log{2}  \nonumber
    \\
    & \hspace{-1cm} -\frac{5}{16}\pi \Cl_{2}{\left(\frac{\pi}{3}\right)} \log{2} + \frac{9}{8} \Ree{\left[G{\left(e^{-\frac{i\pi}{3}},1,-1;1\right)}\right]}\log{2} +\frac{167}{48}\log^2{2}+\frac{503}{2400}\pi^2 \log^2{2} \nonumber
    \\
    & \hspace{-1cm}-\frac{55}{128}\log^3{2} + \frac{55}{64} \pi^2 \log{3} -\frac{3}{32}\pi \Cl_{2}{\left(\frac{\pi}{3}\right)}\log{3} +\frac{55}{64}\log^2{2}\log{3}+ \frac{9}{16}\pi^2 \Li_{2}{\left(-\frac{1}{2}\right)} \nonumber
    \\
    & \hspace{-1cm} -\frac{55}{64}\Li_{2}{\left(-\frac{1}{2}\right)}\log{2}-\frac{91883}{5760}\zeta_3-\frac{607}{480}\zeta_3 \log{2} + i \pi \left[-\frac{5}{2}\log{2}+\frac{1}{8}\pi^2 \log{2}\right] \nonumber
    \\
    & \hspace{-1cm} +\frac{3}{2}F_{3}^{(0)} +\frac{1}{4}F_{12}^{(1)}+\frac{9}{2}F_{14}^{(0)}-\frac{49}{600}F_{15}^{(0)}+\frac{8}{3}F_{18}^{(0)}-\frac{1}{4}F_{18}^{(1)}+F_{22}^{(0)}-\frac{249}{200}F_{31}^{(0)}-\frac{94}{15}F_{57}^{(0)} \nonumber
    \\
    & \hspace{-1cm}+\frac{3}{20}F_{57}^{(1)} +\frac{83}{72}F_{64}^{(0)}-\frac{59}{36}F_{65}^{(0)},
    \\[12pt]
    a_{4}^{(2)}=& -\frac{4261}{144} +\frac{1199}{432}\pi^2 +\frac{90767}{1536000}\pi^4-\frac{185}{576}\pi\Cl_{2}{\left(\frac{\pi}{3}\right)}-\frac{9}{64} \pi \Ime{\left[G{\left(0,1,e^{-\frac{2i\pi}{3}};1\right)}\right]} \nonumber
    \\
    &\hspace{-1cm} -\frac{3}{16}\pi\Ime{\left[G{\left(0,e^{-\frac{i\pi}{3}},-1;1\right)}\right]} -\frac{37}{32}\Ree{\left[G{\left(e^{-\frac{i\pi}{3}},1,-1;1\right)}\right]} +\frac{6843}{3200}\Ree{\left[G{\left(0,0,e^{-\frac{i\pi}{3}},-1;1\right)}\right]} \nonumber
    \\
    &\hspace{-1cm} +\frac{20529}{12800}\Ree{\left[G{\left(0,0,e^{-\frac{2i\pi}{3}},1;1\right)}\right]} +\frac{9}{8}\Ree{\left[G{\left(e^{-\frac{i\pi}{3}},1,1,-1;1\right)}\right]}+\frac{65}{24}\log{2}-\frac{925}{384}\pi^2 \log{2} \nonumber
    \\
    &\hspace{-1cm} - \frac{5}{32}\pi \Cl_{2}{\left(\frac{\pi}{3}\right)}\log{2}+\frac{9}{16}\Ree{\left[G{\left(e^{-\frac{i\pi}{3}},1,-1;1\right)}\right]}\log{2} +\frac{27}{16}\log^2{2}+\frac{1357}{4800}\pi^2 \log^2{2} \nonumber
    \\
    &\hspace{-1cm} -\frac{37}{128}\log^3{2}+\frac{37}{64}\pi^2 \log{3}-\frac{3}{64}\pi \Cl_{2}{\left(\frac{\pi}{3}\right)}\log{3} +\frac{37}{64}\log^2{2} \log{3} +\frac{9}{32}\pi^2 \Li_2{\left(-\frac{1}{2}\right)} \nonumber
    \\
    &\hspace{-1cm} -\frac{37}{64}\Li_2{\left(-\frac{1}{2}\right)}\log{2}-\frac{1849}{1152}\zeta_3-\frac{3953}{960}\zeta_3 \log{2}+F_{3}^{(0)}-\frac{1}{8}F_{12}^{(1)}+3F_{14}^{(0)}-\frac{253}{600}F_{15}^{(0)} \nonumber
    \\
    &\hspace{-1cm} +\frac{15}{8}F_{18}^{(0)}+\frac{1}{8}F_{18}^{(1)} +\frac{5}{4}F_{22}^{(0)}-\frac{41}{25}F_{31}^{(0)}-\frac{57}{20}F_{57}^{(0)}-\frac{3}{40}F_{57}^{(1)}+\frac{77}{36}F_{64}^{(0)}-\frac{367}{144}F_{65}^{(0)},
	\\[12pt]
    a_{5}^{(2)}=& \frac{28525}{5184} +\frac{11}{32}\pi^2 -\frac{13937321}{248832000} \pi^4 -\frac{5}{288}\pi Cl_{2}{\left(\frac{\pi}{3}\right)}+\frac{23}{128}\pi \Ime{\left[G{\left(0,1,e^{-\frac{2i\pi}{3}};1\right)}\right]} \nonumber
    \\
    &\hspace{-1cm} +\frac{23}{96}\pi \Ime{\left[G{\left(0,e^{-\frac{i\pi}{3}},-1;1\right)}\right]}-\frac{1}{16} \Ree{\left[G{\left(e^{-\frac{i\pi}{3}},1,-1;1\right)}\right]}-\frac{20789}{6400}\Ree{\left[G{\left(0,0,e^{-\frac{i\pi}{3}},-1;1\right)}\right]} \nonumber
    \\
    &\hspace{-1cm} -\frac{62367}{25600}\Ree{\left[G{\left(0,0,e^{-\frac{2i\pi}{3}},1;1\right)}\right]} -\frac{23}{16}\Ree{\left[G{\left(e^{-\frac{i\pi}{3}},1,1,-1;1\right)}\right]}-\frac{17}{27}\log{2}-\frac{547}{576}\pi^2 \log{2} \nonumber
    \\
    &\hspace{-1cm} -\frac{5}{128}\pi^3 \log{2} + \frac{115}{576}\pi \Cl_{2}{\left(\frac{\pi}{3}\right)}\log{2} -\frac{23}{32} \Ree{\left[G{\left(e^{-\frac{i\pi}{3}},1,-1;1\right)}\right]}\log{2}-\frac{331}{144}\log^2{2} \nonumber
    \\
    &\hspace{-1cm} +\frac{1}{2}C \log^2{2}-\frac{2287}{3200}\pi^2 \log^2{2}+\frac{343}{576}\log^3{2}-\frac{3}{32}\pi \log^3{2}+\frac{1}{4}\log^4{2}+\frac{1}{32}\pi^2 \log{3} \nonumber
    \\
    &\hspace{-1cm} +\frac{23}{384}\pi \Cl_2{\left(\frac{\pi}{3}\right)}\log{3}+\frac{1}{32}\log^2{2}\log{3}+\frac{7}{6144} \psi^{3}{\left(\frac{1}{4}\right)}-\frac{1}{2048}\psi^{3}{\left(\frac{3}{4}\right)} -\frac{23}{64}\pi^2 \Li_2{\left(-\frac{1}{2}\right)} \nonumber
    \\
    &\hspace{-1cm} -\frac{1}{32}\Li_2{\left(-\frac{1}{2}\right)}\log{2}+\frac{2041}{384}\zeta_3+\frac{7}{64}\pi \zeta_3 -\frac{12023}{5760}\zeta_3 \log{2}+\frac{i}{8} \hpli{\left(0,-,+,-\right)} \nonumber
    \\
    &\hspace{-1cm} + i \pi \left[\frac{43}{108} -\frac{5}{72}\pi^2 +\frac{85}{36}\log{2}+\frac{1}{3}\pi^2 \log{2}-\frac{11}{12}\log^2{2}-\log^3{2} +\zeta_3 \right] \nonumber
    \\
    &\hspace{-1cm} + \frac{3i}{2} \log{2} \left(G{\left(0,0,1;\frac{1}{2}+\frac{i}{2}\right)}-G{\left(0,0,1;\frac{1}{2}-\frac{i}{2}\right)}\right) -\frac{1}{2}F_{3}^{(0)}-\frac{1}{8}F_{12}^{(1)}-\frac{3}{2}F_{14}^{(0)}-\frac{7}{150}F_{15}^{(0)} \nonumber
    \\
    &\hspace{-1cm} -\frac{7}{8}F_{18}^{(0)}+\frac{1}{8}F_{18}^{(1)}-\frac{1}{4}F_{22}^{(0)}+\frac{47}{200}F_{31}^{(0)}+\frac{43}{20}F_{57}^{(0)}-\frac{3}{40}F_{57}^{(1)}-\frac{3}{8}F_{64}^{(0)}+\frac{9}{16}F_{65}^{(0)},
    \\[12pt]
    a_{6}^{(2)}=& \frac{100987}{10368} -\frac{1451}{1728}\pi^2 -\frac{3574663}{165888000}\pi^4 +\frac{325}{2304}\pi \Cl_2{\left(\frac{\pi}{3}\right)}+\frac{3}{256}\pi \Ime{\left[G{\left(0,1,e^{-\frac{2i\pi}{3}};1\right)}\right]}  \nonumber
    \\
    &\hspace{-1cm} +\frac{1}{64}\pi \Ime{\left[G{\left(0,e^{-\frac{i\pi}{3}},-1;1\right)}\right]}+\frac{65}{128}\Ree{\left[G{\left(e^{-\frac{i\pi}{3}},1,-1;1\right)}\right]} + \frac{8199}{12800} \Ree{\left[G{\left(0,0,e^{-\frac{i\pi}{3}},-1;1\right)}\right]} \nonumber
    \\
    &\hspace{-1cm} +\frac{24597}{51200} \Ree{\left[G{\left(0,0,e^{-\frac{2i\pi}{3}},1;1\right)}\right]}-\frac{3}{32}\Ree{\left[G{\left(e^{-\frac{i\pi}{3}},1,1-1;1\right)}\right]}+\frac{289}{864}\log{2}+\frac{35143}{23040}\pi^2 \log{2} \nonumber
    \\
    &\hspace{-1cm} +\frac{5}{256}\pi^3 \log{2} +\frac{5}{384}\pi \Cl_2{\left(\frac{\pi}{3}\right)} \log{2} -\frac{3}{64} \Ree{\left[G{\left(e^{-\frac{i\pi}{3}},1,-1;1\right)}\right]}\log{2}-\frac{737}{576}\log^2{2}-\frac{1}{4}C \log^2{2} \nonumber
    \\
    &\hspace{-1cm} -\frac{1049}{19200}\pi^2 \log^2{2}+\frac{841}{4608}\log^3{2}+\frac{3}{64}\pi \log^3{2} +\frac{1}{8}\log^4{2}-\frac{65}{256}\pi^2 \log{3} +\frac{1}{256}\pi \Cl_2{\left(\frac{\pi}{3}\right)}\log{3} \nonumber
    \\
    &\hspace{-1cm} -\frac{65}{256}\log^2{2}\log{3}-\frac{7}{12288}\psi^3{\left(\frac{1}{4}\right)} + \frac{1}{4096}\psi^3{\left(\frac{3}{4}\right)} -\frac{3}{128}\pi^2 \Li_2{\left(-\frac{1}{2}\right)}+\frac{65}{256}\Li_2{\left(-\frac{1}{2}\right)}\log{2} \nonumber
    \\
    &\hspace{-1cm} +\frac{16553}{7680}\zeta_3 -\frac{7}{128}\pi \zeta_3 +\frac{1517}{1280}\zeta_3 \log{2}+\frac{3i}{4} \log{2}\left(G{\left(0,0,1;\frac{1}{2}-\frac{i}{2}\right)}-G{\left(0,0,1;\frac{1}{2}+\frac{i}{2}\right)}\right) \nonumber
    \\
    &\hspace{-1cm} - \frac{i}{16}\hpli{\left(0,-,+,-\right)} - \frac{1}{4}F_3^{(0)}+\frac{1}{16}F_{12}^{(1)}-\frac{3}{4}F_{14}^{(0)}+\frac{37}{300}F_{15}^{(0)} -\frac{23}{48}F_{18}^{(0)}-\frac{1}{16}F_{18}^{(1)}-\frac{3}{8}F_{22}^{(0)} \nonumber
    \\
    &\hspace{-1cm} +\frac{173}{400}F_{31}^{(0)}+\frac{53}{120}F_{57}^{(0)}+\frac{3}{80}F_{57}^{(1)}-\frac{125}{144}F_{64}^{(0)}+\frac{293}{288}F_{65}^{(0)},
    \\[12pt]
    a_{7}^{(2)}=& \frac{32783}{2592} -\frac{2713}{3456}\pi^2 -\frac{682801}{49766400}\pi^4 +\frac{245}{1152}\pi \Cl_2{\left(\frac{\pi}{3}\right)}+\frac{11}{128}\pi \Ime{\left[G{\left(0,1,e^{-\frac{2i\pi}{3}};1\right)}\right]}  \nonumber
    \\
    &\hspace{-1cm} +\frac{11}{96}\pi \Ime{\left[G{\left(0,e^{-\frac{i\pi}{3}},-1;1\right)}\right]} +\frac{49}{64}\Ree{\left[G{\left(e^{-\frac{i\pi}{3}},1,-1;1\right)}\right]} -\frac{4909}{1280}\Ree{\left[G{\left(0,0,e^{-\frac{i\pi}{3}},-1;1\right)}\right]} \nonumber
    \\
    &\hspace{-1cm} -\frac{14727}{5120} \Ree{\left[G{\left(0,0,e^{-\frac{2i\pi}{3}},1;1\right)}\right]}-\frac{11}{16}\Ree{\left[G{\left(e^{-\frac{i\pi}{3}},1,1,-1;1\right)}\right]} +\frac{65}{216}\log{2}+\frac{15979}{11520}\pi^2\log{2} \nonumber
    \\
    &\hspace{-1cm} +\frac{55}{576}\pi \Cl_2{\left(\frac{\pi}{3}\right)} \log{2} -\frac{11}{32} \Ree{\left[G{\left(e^{-\frac{i\pi}{3}},1,-1;1\right)}\right]}\log{2}-\frac{803}{288}\log^2{2} -\frac{301}{1920}\pi^2 \log^2{2} \nonumber
    \\
    &\hspace{-1cm}+\frac{697}{2304}\log^3{2}+\frac{3}{8}\log^4{2}-\frac{49}{128}\pi^2 \log{3}+\frac{11}{384}\pi \Cl_2{\left(\frac{\pi}{3}\right)} \log{3} -\frac{49}{128}\log^2{2}\log{3} \nonumber
    \\
    &\hspace{-1cm} -\frac{11}{64}\pi^2 \Li_2{\left(-\frac{1}{2}\right)}+\frac{49}{128}\Li_2{\left(-\frac{1}{2}\right)} \log{2}-3\Li_4{\left(\frac{1}{2}\right)}+\frac{43837}{11520}\zeta_3 -\frac{1723}{1152}\zeta_3 \log{2} \nonumber
    \\
    &\hspace{-1cm} i\pi \left[\frac{77}{432}+\frac{103}{72}\log{2}+\frac{1}{96}\pi^2 \log{2}-\frac{5}{24}\log^2{2}-\frac{1}{2}\log^3{2}+\frac{3}{16}\zeta_3\right] -\frac{1}{2}F_{3}^{(0)}-\frac{3}{2}F_{14}^{(0)} \nonumber
    \\
    &\hspace{-1cm} +\frac{1}{10}F_{15}^{(0)}-\frac{11}{12}F_{18}^{(0)}-\frac{1}{2}F_{22}^{(0)}+\frac{11}{20}F_{31}^{(0)}+\frac{91}{60}F_{57}^{(0)}-\frac{19}{18}F_{64}^{(0)}+\frac{187}{144}F_{65}^{(0)}.
\end{align}

We now collect the coefficients $b_i^{(2)}$ needed for the light-by-light contributions,
\begin{align}
    b_{1}^{(2)}=&\left[-\frac{631}{24}+\frac{7}{12}\pi^2+\frac{2}{3}\log{2}+\frac{113}{60}\pi^2\log{2}+\frac{2}{3}\pi^2\log{\left(-1+\sqrt{2}\right)}-\frac{4}{3}\log^3{\left(-1+\sqrt{2}\right)}\right.
    \nonumber\\
    &\left. -2\Li_3{\left(3-2\sqrt{2}\right)}-\frac{647}{120}\zeta_3\right]+i \pi\left[-\frac{1}{3}+\frac{1}{4}\pi^2-2\log^2{\left(-1+\sqrt{2}\right)}\right]
    \nonumber\\
    & -6 F_{14}^{(0)}+\frac{2}{3} F_{24}^{(0)}+\frac{2}{5}F_{57}^{(0)}+\frac{7}{3} F_{64}^{(0)}-3 F_{65}^{(0)},
    \\[12pt]
    b_{2}^{(2)}=&\left[-\frac{\pi^2}{6}+2\log{2}+\frac{1}{6}\pi^2\log{2}-\frac{1}{8}\zeta_3\right]+i \pi\left[-\frac{5}{3}+\frac{1}{9}\pi^2\right],
    \\[12pt]
    b_{3}^{(2)}=& -\frac{91}{24}+\frac{1}{6}\pi^2 + \frac{1}{4} G{\left(0,0,1;3-2\sqrt{2}\right)} - \frac{1}{3}\log{2} - \frac{7}{6}\pi^2 \log{2} - \frac{1}{4}\pi^2 \log{\left(-1+\sqrt{2}\right)} \nonumber
    \\
    &\hspace{-1cm} + \frac{1}{2}\log^3{\left(-1+\sqrt{2}\right)} + \Li_3{\left(3-2\sqrt{2}\right)} + \frac{137}{24}\zeta_3 + i \pi \left[\frac{1}{6}+\frac{3}{4}\log^2{\left(-1+\sqrt{2}\right)}\right] \nonumber
    \\
    &\hspace{-1cm} +\sqrt{2}\left[-\frac{\pi^2}{12}-\frac{1}{2}G{\left(0,1;3-2\sqrt{2}\right)}+\frac{1}{2}\log^2{\left(-1+\sqrt{2}\right)}+\frac{1}{2}i\pi\log{\left(-1+\sqrt{2}\right)}\right] \nonumber
    \\
    &\hspace{-1cm}+ 3 F_{14}^{(0)}-\frac{1}{3} F_{24}^{(0)}+F_{57}^{(0)}+\frac{1}{3} F_{64}^{(0)}-\frac{1}{2} F_{65}^{(0)}.
	\\[12pt]
    b_{4}^{(2)}=& -\frac{5}{48}\pi^2 +\frac{3}{2}\log{2}+\frac{1}{6}\pi^2 \log{2}-\frac{19}{16}\zeta_3 + i \pi \left[-\frac{11}{12}+\frac{1}{9}\pi^2\right].
\end{align}

In the following, we list the $c_i^{(2)}$ coefficients needed for the vaccuum polarisation contributions,
\begin{align}
    c_{1}^{(2)}=&-\frac{15061}{1440}-\frac{107}{2160}\pi^2-\frac{1}{60}\pi^2 \log{2}+\frac{8}{15}\zeta_3+\frac{1}{5} F_{57}^{(0)}+\frac{217}{180} F_{64}^{(0)}-\frac{19}{9} F_{65}^{(0)},
    \\[12pt]
    c_{2}^{(2)}=&\frac{41}{36}-\frac{13}{144}\pi^2-\frac{2}{3}\log{2}-\frac{7}{24}\zeta_3,
    \\[12pt]
    c_{3}^{(2)}=& \frac{38603}{2592}-\frac{5}{27}\pi^2-\frac{649}{691200}\pi^4-\frac{2}{9}G{\left(0,0,1;3-2\sqrt{2}\right)} + \frac{21}{160} \Ree{\left[G{\left(0,0,e^{-{i\pi}{3}},-1;1\right)}\right]} \nonumber
    \\
    &\hspace{-1cm} + \frac{63}{640} \Ree{\left[G{\left(0,0,e^{-{2i\pi}{3}},1;1\right)}\right]} - \frac{11}{108}\log{2} + \frac{349}{240}\pi^2 \log{2} - \frac{11}{240}\pi^2 \log^2{2} + \frac{5}{18}\pi^2 \log{\left(-1+\sqrt{2}\right)} \nonumber
    \\
    &\hspace{-1cm} - \frac{5}{9} \log^3{\left(-1+\sqrt{2}\right)} - \frac{19}{18} \Li_3{\left(3-2\sqrt{2}\right)} - \frac{46}{5} \zeta_3 + \frac{7}{48} \log{2} \zeta_3 + i\pi \left[\frac{1}{108} - \frac{5}{6}\log^2{\left(-1+\sqrt{2}\right)}\right] \nonumber
    \\
    &\hspace{-1cm} + \sqrt{2} \left[\frac{2}{27}\pi^2 + \frac{4}{9} G{\left(0,1;3-2\sqrt{2}\right)}-\frac{4}{9}\log^2{\left(-1+\sqrt{2}\right)}-\frac{4}{9}i\pi \log{\left(-1+\sqrt{2}\right)}\right] \nonumber
    \\
    &\hspace{-1cm} + 6 F_{14}^{(0)}+\frac{11}{120} F_{15}^{(0)}+\frac{1}{2} F_{18}^{(0)}+\frac{1}{40} F_{31}^{(0)}-\frac{49}{20} F_{57}^{(0)}-\frac{53}{36} F_{64}^{(0)}+\frac{137}{72} F_{65}^{(0)},
    \\[12pt]
    c_{4}^{(2)}=& - \frac{238901}{25920} + \frac{229}{1440}\pi^2 - \frac{3119}{1382400}\pi^4 +\frac{23}{24}\log{2}-\frac{19}{480}\pi^2 \log{2}-\frac{1}{480}\pi^2 \log^2{2} \nonumber
    \\
    &\hspace{-1cm} +\frac{19}{15}\zeta_3 - \frac{7}{96}\log{2}\zeta_3 + \frac{51}{320}\Ree{\left[G{\left(0,0,e^{-\frac{i\pi}{3}},-1;1\right)}\right]} + \frac{153}{1280}\Ree{\left[G{\left(0,0,e^{-\frac{2i\pi}{3}},1;1\right)}\right]} \nonumber
    \\
    &\hspace{-1cm} -3 F_{14}^{(0)}+\frac{1}{240} F_{15}^{(0)}-\frac{1}{4} F_{18}^{(0)}+\frac{11}{80} F_{31}^{(0)}+\frac{19}{40} F_{57}^{(0)}+\frac{809}{1080} F_{64}^{(0)}-\frac{53}{48} F_{65}^{(0)},
	\\[12pt]
    c_{5}^{(2)}=& -\frac{439}{648} - \frac{41}{27}\log{2} + \frac{1}{18}\pi^2 \log{2} + \frac{5}{9} \log^2{2} - \frac{2}{9}\log^3{2}+\frac{163}{144}\zeta_3 \nonumber
    \\
    &\hspace{-1cm}+ i\pi \left[\frac{25}{27} -\frac{1}{6}\pi^2 - \frac{5}{9}\log{2} + \frac{1}{3}\log^2{2} \right],
    \\[12pt]
    c_{6}^{(2)}=& -\frac{199}{1296} - \frac{11}{54}\log{2}-\frac{1}{72}\pi^2 \log{2} + \frac{4}{9}\log^2{2} - \frac{1}{9}\log^3{2}-\frac{89}{288}\zeta_3.
\end{align}

\bibliographystyle{JHEP}
\bibliography{biblio}

\end{document}